\documentclass[aps,twocolumn,preprintnumbers,amsmath,amssymb,floatfix]{revtex4}

\usepackage{graphicx}
\usepackage{bbm}
\usepackage{bbold}
\usepackage{hyperref}
\usepackage[sort&compress]{natbib}

\newcommand{\vare}{\varepsilon}
\newcommand{\rmi}{{\rm i}}

\begin{document}

\hypersetup{pdftitle={Superconducting quantum criticality in three-dimensional Luttinger semimetals}}
\title{Superconducting quantum criticality in three-dimensional Luttinger semimetals}

\author{Igor Boettcher}
\affiliation{Department of Physics, Simon Fraser University, Burnaby, British Columbia, Canada V5A 1S6}
\author{Igor F. Herbut}
\affiliation{Department of Physics, Simon Fraser University, Burnaby, British Columbia, Canada V5A 1S6}

\begin{abstract}
We study a simple model of three-dimensional fermions close to a quadratic band touching point, built from the celebrated Luttinger single-particle Hamiltonian and an attractive contact interaction between the particles. Such a system displays a novel quantum critical point between the semimetallic and an s-wave superconducting phase at which the low-energy ``Luttinger fermions" are inextricably coupled to the order parameter fluctuations. The quantum critical point is perturbatively accessible near four spatial dimensions,
where it features nontrivial scaling with dynamical exponent $1<z<2$ and emergent rotational and particle-hole symmetries. Some features of the criticality, such as oscillatory corrections to scaling and its enhanced symmetry, are discussed.

\end{abstract}

%\pacs{74.40.Kb, 05.10.Cc}
%05.10.Cc Renormalization group methods
%74.40.Kb Quantum critical phenomena (in 74. Superconductivity)

\maketitle

\section{Introduction}\label{SecIntro}

Materials with Fermi points such as Dirac or Weyl semimetals have attracted immense interest recently as the modified electronic dispersion provides a portal to the fascinating phenomenology of quantum condensed matter. From the point of view of electron-electron correlations, three-dimensional (3D) systems with a quadratic band touching (QBT) point could be particularly interesting. For example, it has been pointed out long ago by Abrikosov \cite{abrikosov}, and reexamined closely more recently \cite{moon}, that the long-range nature of the Coulomb interaction may cause the ground state of such a system to be possibly the simplest instance of a non-Fermi liquid. The required dispersion arises naturally in 3D gapless semiconductors in which spin-orbit coupling is strong enough so that band inversion causes the Fermi level to lie at a QBT point \cite{abrben}, famously exemplified by gray tin or mercury telluride. As the low-energy electronic degrees of freedom in these systems are described by the Luttinger Hamiltonian \cite{luttinger,murakami}, one might call them ``Luttinger semimetals''.

Due to the generation of a short-range part of repulsive interactions by the long-range Coulomb forces, it has been argued that Abrikosov's  non-Fermi liquid may be unstable towards a gapped nematic phase, in which rotational symmetry is spontaneously broken \cite{herbut2014, janssen2015}. In particular, the structure of the 3D Luttinger Hamiltonian results in a certain frustration which complicates lowering the ground state energy by gapping out the ``Luttinger fermions'': Indeed, the rotationally symmetric excitonic (particle-hole) gap is not even available, and the next best option for the reduction of the energy of the filled Fermi sea is to open the cylindrically symmetric gap, which transforms as a second-rank irreducible tensor under rotations. Novel magnetic instabilities in the iridates, believed to have an equivalent low-energy quasiparticle spectrum \cite{rhim,kondo}, have also been proposed and studied \cite{savary,murray}.

From the point of view of the theory of phase transitions, quantum criticality in the 3D QBT system would present an interesting and novel example in which low-energy fermions are coupled to the order parameter, but within a field theory that has at most Galilean, and not Lorentz invariance. One may wonder, for example, whether the Galilean invariance \cite{leblond, hagen}, similarly to the Lorentz case, also imposes some constraints on universal quantities at the transition. For example, the dynamical critical exponent $z$ is fixed to unity in the Lorentz case \cite{herbut2006, herjurroy, assaad, parisen, sorella}. Some of these issues have been studied for the aforementioned nematic quantum critical point \cite{janssen}. In that case, one can show that for the nematic transition $z$ remains equal to two to the leading order in the appropriately formulated $\vare$-expansion.

The Luttinger Hamiltonian describes a spin-orbit coupled system with total angular momentum $j=3/2$ and a parabolic dispersion. The half-integer nature of the representation of the rotational group dictates that the (unique) anti-unitary operator that provides the time-reversal symmetry (TRS) of the Hamiltonian has a negative square \cite{ballentine}, which, as usual, implies Kramers degeneracy of the spectrum. The same feature of the TRS operator, however, {\it also}  guarantees that the energy spectrum can be gapped in a rotationally invariant way, but in the {\it particle-particle} channel \cite{herbut2013}. The resulting order, which breaks the particle number $U(1)$ symmetry, is nothing but rotationally invariant s-wave superconductivity in the present context of spin-orbit coupled systems. Equivalently, it can be understood as the Majorana mass in the Galilean-invariant fermionic system under consideration \cite{herbut2013}.

Motivated by the growing experimental relevance of 3D materials with QBT, as well as by the potentially interesting theoretical issues discussed above, we consider here a 3D system with the Fermi level at the QBT point and with an attractive, contact, density-density interaction between the Luttinger fermions. One assumes that such a featureless interaction, in the spirit of BCS \cite{bcs}, crudely, may arise from the underlying electron-phonon interaction after the phonons have been integrated out. Since the non-interacting Gaussian fixed point of the theory is attractive, as a consequence of the vanishing density of states at the QBT, the system can have a phase transition only at a finite value of the coupling constant \cite{herbut2014}.
Using the Fierz identities \cite{herjurroy} we show that besides the anticipated s-wave channel, the interaction term is also equally attractive in yet another, unconventional, superconducting channel of the d-wave type, in which five complex order parameters transform under rotations as the components of an irreducible tensor of rank $\ell=2$. This more exotic possibility notwithstanding, we demonstrate that by increasing the coupling constant the zero temperature transition is still into the isotropic s-wave state, which we study in detail. In this work we neglect the repulsive long-range Coulomb interaction between fermions, assuming it to be made sufficiently weak by a large dielectric constant of the lattice.

We construct the minimal field theory for the complex s-wave superconducting order parameter coupled to the Luttinger fermions near the QBT point, and study its universal properties. This is achieved in a controlled manner within a careful generalization of the field theory to $d=4$ spatial dimensions, which is identified as the upper critical dimension \cite{herbutbook}. As a consequence of the dynamical critical exponent $z$ being equal to two at the Gaussian fixed point only the ``Yukawa" coupling between the fermions and the order parameter is relevant in dimensions $2<d<4$. The universal critical properties are governed by a fixed point with an enhanced symmetry, at which, among other things, the dynamical critical exponent is reduced, $z<2$. Both the reduction of the full rotational symmetry down to the cubic symmetry of the lattice and the natural particle-hole asymmetry of the Luttinger Hamiltonian, present away from criticality, are shown to be irrelevant perturbations at the critical point. The former perturbation is found to represent the leading, and remarkably weakly, irrelevant coupling. Particle-hole asymmetry turns out to be particularly interesting as it provides a rare example of oscillatory corrections to scaling.

The paper is organized as follows. In the next section we motivate and define the interacting Hamiltonian we wish to examine. In section \ref{SecQFT} we construct the field theory for the Luttinger fermions coupled to the scalar superconducting order parameter and discuss the structure of the quantum critical point near the upper critical dimension. We provide a broader discussion of our findings in section \ref{SecDisc}. Numerous technical details necessary to derive our results, including the necessary algebra of real Gell-Mann matrices, are presented in three long appendices.

%\enlargethispage{2\baselineskip}

\section{Luttinger fermions with attraction}\label{SecQBTl}

\subsection{Luttinger Hamiltonian}

We are interested in low-energy excitations described by the standard Luttinger Hamiltonian \cite{luttinger} given by
\begin{align}
 \nonumber H= \frac{\hbar^2}{2m}\Bigl[{}&\Bigl(\alpha_1+\frac{5}{2}\alpha_2\Bigr)p^2\mathbb{1}-2\alpha_3(\vec{p}\cdot\vec{J})^2\\
 \label{qbt1} &+2(\alpha_3-\alpha_2)(p_x^2J_x^2+p_y^2J_y^2+p_z^2J_z^2)\Bigr].
\end{align}
Herein, $\vec{p}=-\rmi \nabla$, and $J_i$ are spin $j=3/2$ angular momentum operators, which can be represented by $4\times4$ matrices. The phenomenological Luttinger parameters $\alpha_{1,2,3}$ and the band mass $m$ may be determined experimentally or from numerical simulations for a given material under consideration. The form of $H$ is dictated by $k\cdot p$ perturbation theory and the crystal's cubic symmetry. We assume $|\alpha_1|<2|\alpha_2|$, corresponding to an inverted band structure, and place the Fermi level at $\mu=0$ so that the highest occupied valence band and the (empty) conduction band touch quadratically. For $\alpha_2=\alpha_3$ the Hamiltonian is fully rotationally symmetric, whereas for $\alpha_2\neq\alpha_3$ the rotation symmetry is only cubic.

The same Hamiltonian can also be written in an economical form by using $4\times 4$ euclidean Dirac matrices $\gamma_a$ as
\begin{align}
 \label{qbt2} H = x p^2 + \sum_{a=1}^5 d_a (\vec{p}) \gamma_a + \delta \sum_{a=1} ^5 s_a d_a(\vec{p}) \gamma_a,
\end{align}
with the $\gamma_a$ providing one of the (two possible) irreducible, four-dimensional Hermitian representations of the five-component Clifford algebra defined by the anticommutator $\{ \gamma_a, \gamma_b \} = 2 \delta_{ab}$. The five functions $d_a(\vec{p})$ are the real $\ell=2$ spherical harmonics, given by
\begin{align}
 \nonumber d_1(\vec{p}) &= \frac{\sqrt{3}}{2}(p_x^2-p_y^2),\ d_2(\vec{p}) = \sqrt{3} p_xp_y,\  d_3(\vec{p}) = \sqrt{3} p_xp_z,\\
 \label{qbt2b} d_4(\vec{p}) &= \sqrt{3}p_y p_z,\  d_5(\vec{p}) =\frac{1}{2}(2p_z^2-p_x^2-p_y^2).
\end{align}
For later purpose of generalization to other spatial dimensions, we immediately note that they can be written as
\begin{align}
 \label{qbt3} d_ a (\vec{p}) = \sqrt{ \frac{d}{2 (d-1)} } p_i \Lambda^a_{ij} p_j,
\end{align}
where $d=3$ and $\Lambda^a$ are the five real Gell-Mann matrices in 3D \cite{janssen}. The parameters $x$ and $\delta$ in Eq. (\ref{qbt2}) are related to the Luttinger parameters by means of $x=\hbar^2\alpha_1/2m$ and $1\mp \delta = -\hbar^2 \alpha_{2,3}/m$. We have chosen units such that $-\hbar^2(\alpha_2+\alpha_3)/2m=1$. The energy spectrum of $H$ for $\delta=0$ is simply given by $(x \pm 1) p^2$. Hence, the parameter $0\leq x<1$ measures the amount of particle-hole asymmetry. For nonzero $\delta$ the full rotational symmetry is reduced to the cubic subgroup thereof. In the anisotropy term proportional to $\delta$, the factor $s_a$ distinguishes between the two diagonal and the three off-diagonal Gell-Mann matrices in the definition in Eq. (\ref{qbt3}).
We choose $s_a=+1$ for the off-diagonal (indices 2,3,4) and $s_a =-1$ for the diagonal (indices 1,5) matrices.

\subsection{Time-reversal symmetry}

The Luttinger Hamiltonian is time-reversal symmetric. One can construct a representation-independent time-reversal operator by
recalling that the five anticommutating $\gamma$-matrices can always be chosen such that three are real and two are imaginary \cite{herbut2012}. The reader may understand this  as a generalization of the more familiar feature of the Pauli matrices, where two are real and one is imaginary. Let us choose a representation in which the two imaginary ones are $\gamma_4$ and $\gamma_5$. The {\it unique} anti-unitary operator which commutes with $H$ and thus represents the operation of time reversal is then
\begin{align}
 \label{qbt4} T= \gamma_{45} K,
\end{align}
where $\gamma_{45}= \rmi \gamma_4 \gamma_5 $ and $K$ is complex conjugation. Evidently, $T^2 = -1$, and the spectrum of $H$ is doubly (Kramers) degenerate. This expresses the fact that the four-dimensional spinor representation of $ SO(5)$, generated by the ten generators $\gamma_{ab}=\rmi \gamma_a\gamma_b$, $a<b$,  is {\it pseudoreal} \cite{georgi}. This, of course, also agrees with the Luttinger Hamiltonian describing particles of half-integer spin \cite{ballentine}.

\subsection{Attractive interactions}

We now define the euclidean zero temperature quantum mechanical action by
\begin{align}
 \label{qbt5} \mathcal{S}[\psi] = \int \mbox{d}\tau \mbox{d}^dx \Bigl( \psi^\dagger (\partial_\tau +H)\psi + 4 u (\psi^\dagger \psi)^2 \Bigr),
\end{align}
with imaginary time $\tau$, attractive coupling constant $u<0$, and four-component Grassmann field $\psi(\tau,\vec{x})=(\psi_1,\psi_2,\psi_3,\psi_4)^{\rm T}$.
By perturbative power counting we deduce $u \sim (\text{length})^{d+z-4}$, so that for a quadratic dispersion with $z=2$, a sufficiently weak coupling $u$ is irrelevant for $d=3$. There is therefore no usual Cooper instability for infinitesimal $u$, essentially because the density of states vanishes at the QBT \cite{nozieres}. Nevertheless, a quantum phase transition is still expected to occur at sufficiently large coupling, where the system may lower its ground state energy by opening a gap at the Fermi level. Furthermore, using the Fierz identity (\ref{fierz13}) one can show that
 \begin{align}
\label{qbt6} (\psi^\dagger\psi)^2 = \frac{1}{4}(L_{\rm s}+L_{\rm d})
\end{align}
with
\begin{align}
 \label{qbt7} L_{\rm s} &= (\psi^\dagger \gamma_{45} \psi^*)(\psi^{\rm T} \gamma_{45} \psi),\\
 \label{qbt8} L_{\rm d} &=\sum_{a=1}^{5} (\psi^\dagger \gamma_a \gamma_{45} \psi^*)(\psi^{\rm T} \gamma_{45} \gamma_a \psi).
\end{align}
The decomposition in Eq. (7)  suggests that there are two competing superconducting orders: $\phi = \langle \psi^{\rm T} \gamma_{45} \psi \rangle$ and $\phi_a = \langle \psi^{\rm T} \gamma_{45} \gamma_a \psi \rangle$. Since $\gamma_{45} \psi^* = T \psi$, and the time reversal operator commutes with all spatial transformations such as rotations, $\phi$ is clearly a scalar. Therefore, $\phi$ is the {\it s-wave} superconducting order parameter. For the same reason, the five fields $\phi_a$ transform under rotations the same way as the five gamma matrices $\gamma_a$. Since the term $d_a \gamma_a$ in the Hamiltonian is obviously a scalar, and the $\ell=2$ spherical harmonics $d_a$ transform as the components of an irreducible second-rank tensor, the $\gamma_a$ do so as well \cite{herbut2014, janssen}. Hence, the five components of the order parameters $\phi_a$ transform under rotations like the $\ell=2$ spherical harmonics, and thus constitute a {\it d-wave} order parameter.

It is easy to show, however, that although the strength of attraction in both s-wave and d-wave channels is the same, s-wave ordering is energetically preferred. Let us introduce Greek indices $\mu,\nu=0,1,\dots,5$ and collect the order parameters in the field $\phi_\mu$, with $\phi_0 = \phi$ and $\phi_{a=1,...5}$ from above. We denote the unit matrix by $\gamma_0=\mathbb{1}$.
The RPA superconducting susceptibility at zero momentum and frequency (Fig. 1b in the text) is then 
\begin{equation}
 \chi^{-1} _{\mu\nu} = \frac{\delta_{\mu\nu}}{|u|} - \frac{1}{2} \Bigl( \delta_{\mu\nu} + \frac{1}{20} \mbox{tr} (\gamma_b \gamma_\mu \gamma_b \gamma_\nu)\Bigr)\int \frac{\mbox{d}^d q}{(2\pi)^d}\frac{1}{q^2},
 \end{equation}
where we have set the parameters $x=\delta=0$ for simplicity, and there is an upper cutoff in the momentum integral. Computing the trace gives
\begin{equation}
 \chi^{-1} _{\mu\nu} =  \delta_{\mu\nu} \Bigl(\frac{1}{|u|}- C_\mu \int \frac{\mbox{d}^d q}{(2\pi)^d}\frac{1}{q^2} \Bigr)
 \end{equation}
 with $C_0=1$ and $C_a=1/5$. With the increase of $|u|$ the first to diverge is therefore the s-wave susceptibility at a finite (non-universal) $u_c$, whereas the d-wave susceptibility would, in the absence of the s-wave order, diverge only at a higher $u_c ' = 5 u_c$. This shows that there is a wide interval of couplings $|u| > |u_c| $ where the energy of the s-wave superconducting phase is lower than the energy of the normal phase, whereas {\it any} combination of the d-wave order parameters yields an energy still higher than that of the normal phase.

The physical reason is that allowing the s-wave order parameter $\phi \neq 0$ opens up a full rotationally invariant gap, whereas $\phi_a \neq 0$ does not. In more formal terms, only a finite $\phi$ would appear in the mean-field quasiparticle Hamiltonian as a rotational invariant (Majorana) mass term, and as such represents the dominant symmetry breaking channel \cite{herbut2012}.
The ground state energy of the system is therefore reduced more by distributing all of the condensation energy into the s-wave channel, at least near the transition point. Therefore, we set $\phi_a=0$ hereafter and for the time being only consider the s-wave transition.

\section{Quantum critical point}\label{SecQFT}

\subsection{Field theory}

We now formulate the non-relativistic field theory describing the system close to its superconducting quantum phase transition and determine its universal  properties. The corresponding Lagrangian comprises low-energy Luttinger fermions coupled to the fluctuating, complex, bosonic field, whose expectation value represents the superconducting s-wave order parameter. The field theory needs to reflect the cubic symmetry of the lattice, time-reversal invariance, and a global $U(1)$-symmetry due to particle number conservation. The effective Lagrangian has the form
 \begin{align}
 \nonumber L(\psi,\phi) = {}& \psi^\dagger(\partial_\tau+H)\psi + g ( \phi \psi^\dagger \gamma_{45} \psi^* + \phi^* \psi^{\rm T} \gamma_{45} \psi )\\
 \label{qft1} &+\phi^*(y\partial_\tau-c^2\partial_\tau^2-\nabla^2+r)\phi +\lambda|\phi|^4.
\end{align}
The tuning parameter $r$ should be understood as being proportional to $u - u_{\rm c}$, where $u_{\rm c} < 0$ is the critical value of the attractive interaction.
The critical point in this language is thus located at $r=0$. The complex bosonic field $\phi$ is coupled to the fermions precisely as a Majorana mass \cite{herbut2013} and its fluctuations are incorporated by the kinetic term in the second line. We rescaled the fields and the time coordinate so that the coefficients of the terms $|\nabla \phi|^2 $, $\psi^\dagger \partial_\tau \psi$, and $ \psi^\dagger H \psi$ are unity, as discussed in appendix \ref{AppB}. This leaves us with the previously introduced coefficients $x$ and $\delta$, the new coefficients $y$ and $c^2$, and the Yukawa and self-interaction couplings $g$ and $\lambda$, respectively.

Power counting at the non-interacting Gaussian fixed point ($g=\lambda=0$) yields the engineering scaling dimensions
\begin{align}
 \label{qft2} &\text{dim}[g] = \frac{6-d-z}{2},\\
 \label{qft3} &\text{dim}[x] = \text{dim}[\delta] = 0,\\
 \label{qft3b} &\text{dim}[y] = 2-z,\\
 \label{qft4} &\text{dim}[c] = 1-z,\\
 \label{qft5} &\text{dim}[\lambda] = 4-d-z,
\end{align}
with $z=2$. Hence, for $d=4$ the coupling $g$ is marginal, whereas $c$ and the boson self-interaction term $\lambda$ are irrelevant for any dimensions $d > 2$. Therefore, we will study the interacting critical point of the system by generalizing the theory to $d=4-\vare$ spatial dimensions, assuming $0<\vare\ll1$. In this way the universal properties of the quantum phase transition can be captured by retaining only the couplings which are relevant and marginal  at the Gaussian fixed point, i.e., $g,x,y,\delta$ only.

As noted above, for $\delta=0$ the theory acquires full rotational symmetry. For
 $x=y=0$ it also possesses  particle-hole symmetry under the discrete transformation
\begin{align}
 \label{qft6} \psi \to \psi^*,\ \phi &\to \phi^*,\ \gamma_a \to -\gamma_a ^*.
\end{align}
The third mapping changes the sign of an odd number of $\gamma$-matrices, namely the real ones, and thus exchanges the two inequivalent irreducible representations of the five component Clifford algebra \cite{herbut2012}. Due to the enlarged symmetry of the theory at $x=y=\delta=0$, these conditions remain invariant under the renormalization group (RG) transformation which varies the momentum cutoff. Once the critical point of the RG is found, however, one needs to check its stability with respect to small but finite symmetry breaking parameters. As described shortly, we will find them all to be irrelevant perturbations.

Based on the power counting in Eqs. (\ref{qft2})-(\ref{qft5}) we are led to studying the superconducting quantum critical point of the system close to $d=4$ dimensions in terms of the critical ($r=0$) effective Lagrangian
\begin{align}
 \nonumber  L_{\rm c}(\psi,\phi) = {}& \psi^\dagger(\partial_\tau+H)\psi +\phi^*(y\partial_\tau-\nabla^2)\phi\\
 \label{qft7} &+ g ( \phi \psi^\dagger \gamma_{45} \psi^* + \phi^* \psi^{\rm T} \gamma_{45} \psi),
\end{align}
which includes all relevant and marginal couplings at the Gaussian fixed point. The Lagrangian $L_{\rm c}$ constitutes the effective low-energy description of the critical theory after fluctuations with momenta larger than $\Lambda$ have been integrated out. It will feature the following form-invariance property under the scale transformation: When integrating out fluctuations in the momentum interval  $[\Lambda/b,\Lambda]$ in a Wilsonian RG procedure with $b>1$ and at small $\vare=4-d>0$, the effective Lagrangian $L_{\rm c}(b)$ for the momentum modes lower than $\Lambda/b$ will remain in the form of Eq. (\ref{qft7}) when the couplings are replaced by appropriate $b$-dependent running couplings \cite{herbutbook}. This is true at the critical point given by $r=0$. For small $r\neq 0$, however, there will still be a significant crossover range of momenta (or energies, or temperatures), where critical scaling can be observed.

\begin{figure}[t]
\centering
\includegraphics[width=8cm]{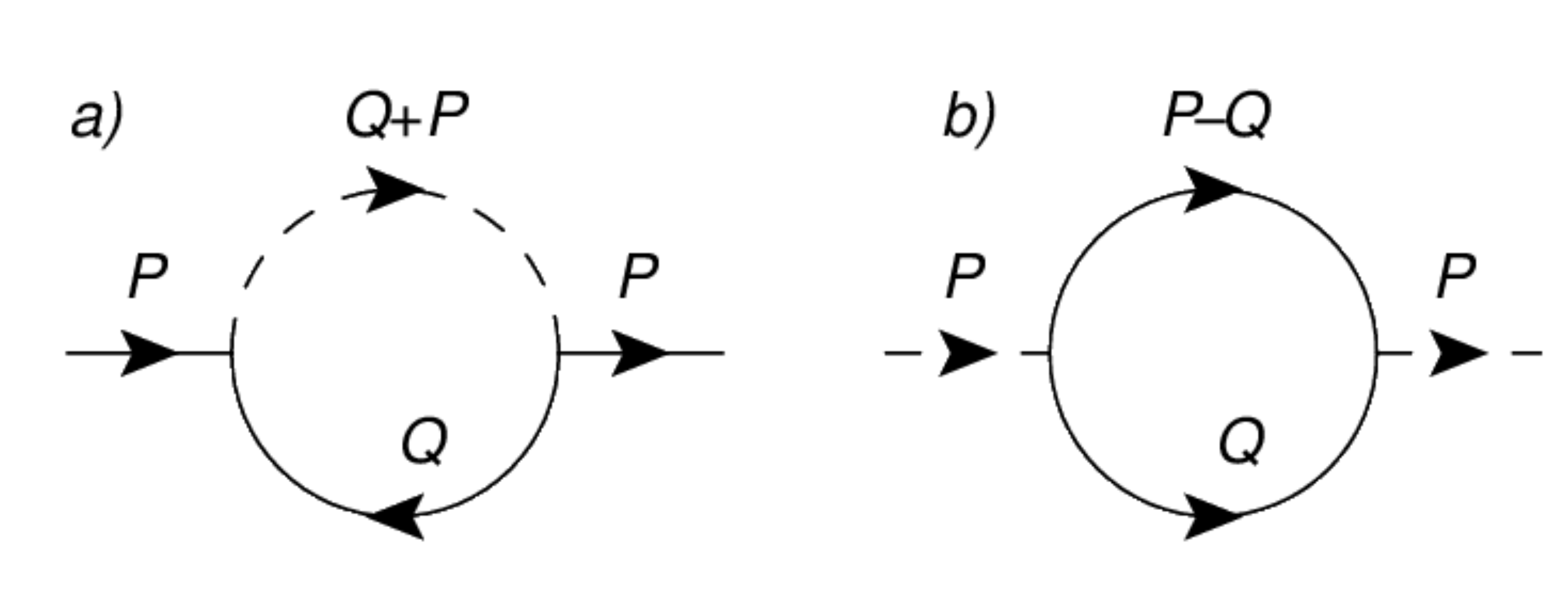}
\caption{Graphic representation of the loops contributing to leading order in the $\vare$-expansion. Panels a) and b) show the diagrams for the fermion self-energy $\Sigma_\psi(P)$ and boson self-energy $\Sigma_\phi(P)$ in Eqs. (\ref{rg7}) and (\ref{rg8}), respectively. A continuous line represents a fermion propagator, a dashed one a boson propagator.}
\label{FigLoops}
\end{figure}

\subsection{RG flow}

\begin{figure}[t]
\centering
\includegraphics[width=4cm]{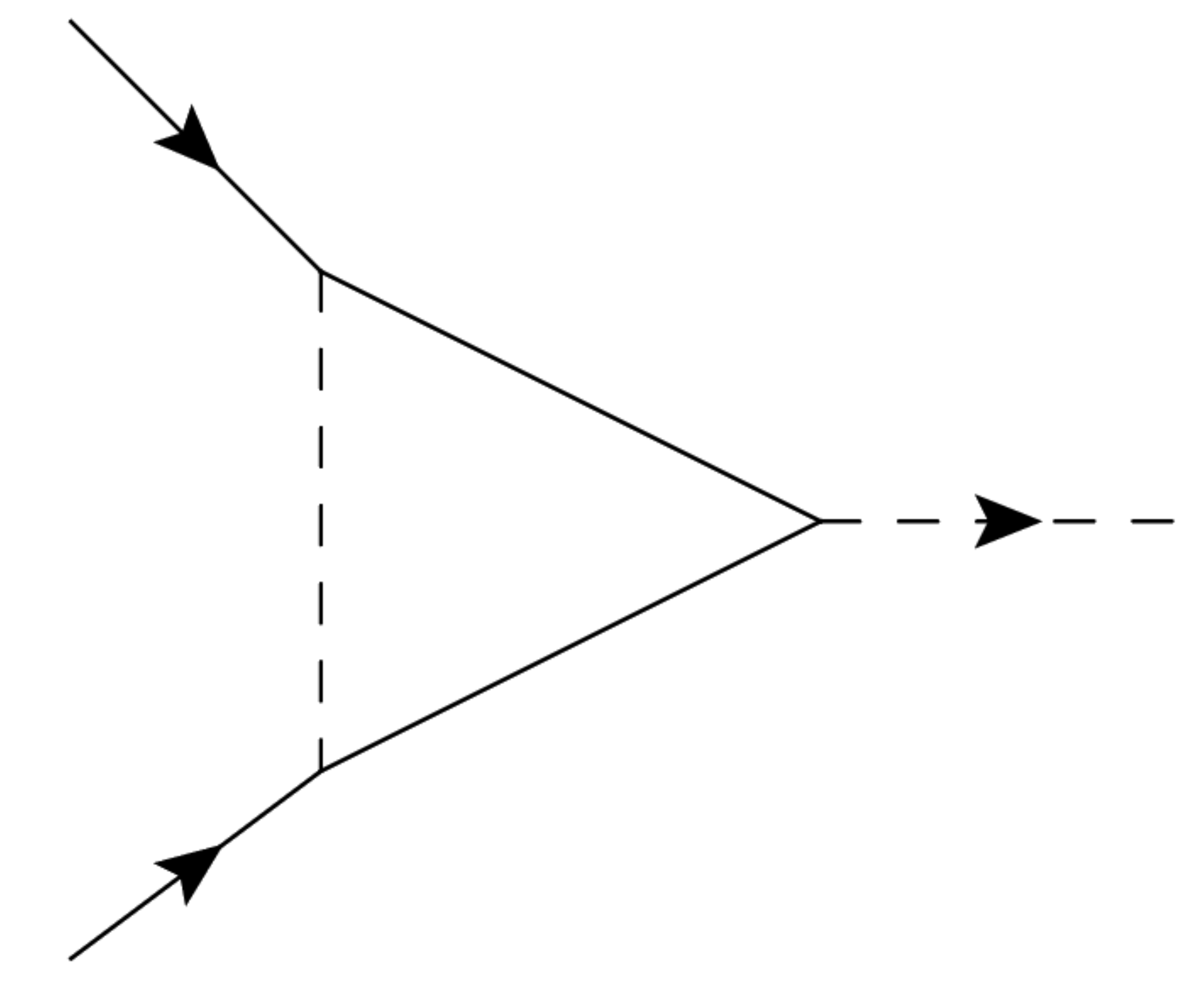}
\caption{One-loop diagram that would be expected to renormalize the Yukawa interaction vertex directly, but which is in fact absent. (The reader can check that no assignment of arrows that would be consistent with the nature of the Yukawa vertex can be made.) }
\label{FigVertex1}
\end{figure}

To the leading, one-loop order in the Yukawa coupling $g$, there are only two (self-energy) diagrams to compute, as depicted in Fig. \ref{FigLoops}. The usual one-loop diagram that would renormalize  the interaction vertex, as in Fig. \ref{FigVertex1}, cannot be formed in the present case, so that the flow of the Yukawa coupling $g$ is solely due to anomalous scaling terms at this order.
The RG evolution of the running couplings defined by Eq. (\ref{qft7}) for small $x,y,\delta$ is then given by
\begin{align}
 \label{qft8} \dot{x} & =-\eta_\psi x +2 y g^2,\\ % + O(g^4, g^2 y^2, g^2 xy),\\
 \label{qft9} \dot{y} &= (2-z-\eta_\phi)y-4 x g^2,\\ % + O(g^4, g^2 x^3),\\
 \label{qft10} \dot{\delta} &= -\eta_\psi \delta +\frac{8}{15}g^2\delta,\\ % + O(g^4 \delta),\\
 \label{qft11} \dot{g} &= \frac{1}{2}(2+\vare-\eta_\phi-2\eta_\psi-z)g, %+ O(g^5),
\end{align}
where the dot denotes a derivative with respect to $\log b$. The anomalous dimensions $\eta_\psi$ and $\eta_\phi$ and the dynamical critical exponent $z$ for small $x,y,\delta$ read
\begin{align}
 \label{qft12} \eta_\psi &= \frac{2}{3}g^2-2yg^2-\frac{8}{45}g^2\delta,\\ % + O(g^4, g^2 y^2),\\
 \label{qft13} \eta_\phi &= 3g^2-g^2\delta,\\ % +O(g^4, g^2 x^2),\\
 \label{qft14} z &= 2-\eta_\psi +2 yg^2. %+O(g^4, g^2 y^2).
\end{align}

The coefficients in the RG equations are computed near $d=4$ spatial dimensions in the following sense. The form of the theory in Eq. (17) is  specific to $d=3$, due to the form of the time reversal operator $T$, which is dimension-dependent. In the perturbation theory we therefore follow the $\gamma$-matrix algebra as in $d=3$, but consider the remaining loop integrals in general dimension. This, for example, requires a generalization of the $\ell=2$ spherical harmonics to dimensions other than $d=3$, but, nevertheless, can be done unambiguously and fully consistent. The reader is invited to consult the appendices about the details of this procedure.

The set of RG flow equations possesses two fixed points, where the right hand sides of the equations vanishes. The first describes the Gaussian fixed point with $g^2=x=y=\delta=0$, unstable for $\vare >0$. At the second, quantum critical fixed point, denoted by a star $\star$, we have
\begin{align}
 \label{qft15} g^2_\star &= \frac{3}{11}\vare,\ x_\star=y_\star=\delta_\star=0.
\end{align}
Since $x,y,\delta$ vanish in both cases, it is consistent to consider the RG equations for small $x,y,\delta$ only. Nevertheless, it is possible to derive more general beta functions, still proportional to $g^2$ but non-perturbative in $x,y$, which we present in appendices \ref{AppRG} and \ref{AppAniso}.

The anomalous scaling behavior at the quantum critical point (QCP) in Eq. (\ref{qft15}) is determined by
\begin{align}
 \label{qft16} \eta_{\psi\star} &= \frac{2}{11}\vare,\ \eta_{\phi\star} = \frac{9}{11}\vare,\ z_\star =2-\frac{2}{11}\vare.
\end{align}
Extrapolating these formulas to $\vare =1$, i.e., $d=3$, one can estimate that
\begin{align}
 \label{qft17} \eta_{\psi\star} &= 0.18,\ \eta_{\phi\star} = 0.82,\ z_\star =1.82.
\end{align}
We observe that the fermion anomalous dimension, and the consequent deviation of $z$ from two are, albeit finite, reasonably small. The boson anomalous dimension, on the other hand, is rather large, since it is proportional to the number of fermionic components, which here is four. This is a common property of this Yukawa-type of theories \cite{assaad, parisen, sorella, herjurvaf,PhysRevB.87.041401, janssen2014,PhysRevA.89.053630,wang2014, li, wang2015, huffman}.

The corresponding anomalous scaling of observables can be measured  in frequency- and momentum-resolved experiments such as, for instance, determinations of the spectral function from ARPES. The dynamical critical exponent $z\neq 2$ will also influence the temperature dependence of several observables, such as the specific heat at the critical point, and the frequency dependence of the conductivity. The nontrivial behavior of the fermionic dispersion, encoded in $\eta_\psi$ and $z$, reflects the fact that the discovered QCP describes fermionic quantum matter without classical analogue.

\subsection{Corrections to scaling}

In order to determine the stability of the QCP we compute the eigenvalues $\{\theta_k\}$ of the stability matrix $M$ with matrix elements
\begin{align}
 \label{qft18} M_{ij}  = \frac{\partial \beta_{G_i}}{\partial G_j}\Bigr|_{G_\star},
\end{align}
where $\beta_{G_i}$ is the beta function of the coupling $G_i\in\{r,g,x,y,\delta\}$. A stable fixed point is characterized by having at most one relevant direction, defined through an eigenvalue $\theta_k$ with positive real part. One relevant direction in coupling space for both the Gaussian fixed point and the QCP consists in $r$, corresponding to the fine-tuning
which is required to ensure $r=0$. At the Gaussian fixed point, the eigenvalues of $M$ are given by the perturbative scaling dimensions of the coupling in Eqs. (\ref{qft2})-(\ref{qft3b}) and thus read $\{\theta_k\}=(2,\vare/2,0,0,0)$ such that the Gaussian fixed point is not stable.

We now show that, in contrast, the interacting QCP is stable. For this purpose we first discuss the relevant direction, $r$, and then explore the interesting scaling properties of the irrelevant couplings. From Eq. (\ref{rg25b}) we deduce the RG flow of $r$ close to the QCP to be
\begin{align}
 \label{rg32} \dot{r} = (2-\eta_\phi)r -4g^2+ O(g^4).
\end{align}
For $\log b\to \infty$ we define the correlation length exponent $\nu$ such that $r(b) \sim r(1) b^{1/\nu}$. If $\nu>0$, we need to fine-tune $r(1)$ for the critical theory such that this expression remains finite for all $b$. Hence we have
\begin{align}
 \label{rg33} \frac{1}{\nu} = \frac{\partial \beta_r}{\partial r}\Bigr|_\star = (2-\eta_{\phi\star}) = 2-\frac{9}{11}\vare+O(\vare^2).
\end{align}
Equivalently,
\begin{align}
 \label{rg34} \nu = \frac{1}{2}+\frac{9}{44}\vare+O(\vare^2).
\end{align}
The susceptibility exponent $\gamma$, which satisfies the scaling relation $\gamma=\nu(2-\eta_\phi)$, is therefore found to be $\gamma_\star =1+ O(\epsilon^2)$ at the QCP, i.e. unchanged from the mean-field value to this order in calculation.

The scaling behavior of the irrelevant couplings $\{g-g_\star,x,y,\delta\}$ close to the critical point, which decouples from that of $r$, is given by the eigenvalues
\begin{align}
 \label{qft19} \theta_\delta & = -\frac{2}{55}\vare,\\
 \label{qft20} \theta_{1} &= \Bigl(-\frac{9}{22} +\rmi \frac{\sqrt{263}}{22}\Bigr)\vare,\\
 \label{qft21} \theta_{2} &= \Bigl(-\frac{9}{22} -\rmi \frac{\sqrt{263}}{22}\Bigr)\vare,\\
 \label{qft22} \theta_{g} &= -\vare.\
\end{align}
Indeed, we observe all real parts to be negative for $\vare>0$. We find the leading correction to scaling to be governed by the parameter $\delta$, i.e., the reduction of full rotation symmetry to cubic symmetry. As the corresponding eigenvalue $-(2/55)\vare$ is exceptionally small, the leading correction to scaling needs to be incorporated in any attempt to observe scaling for a realization of the QBT system on a cubic lattice.

\begin{figure}[t!]
\centering
\begin{minipage}{0.46\textwidth}
\includegraphics[width=7.0cm]{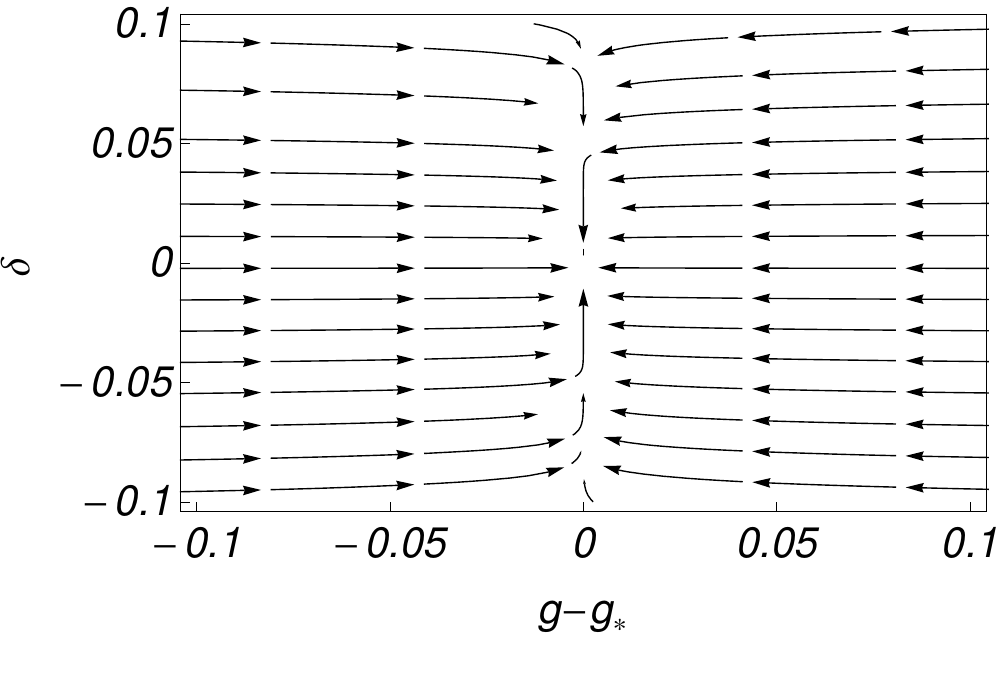}
\includegraphics[width=7.0cm]{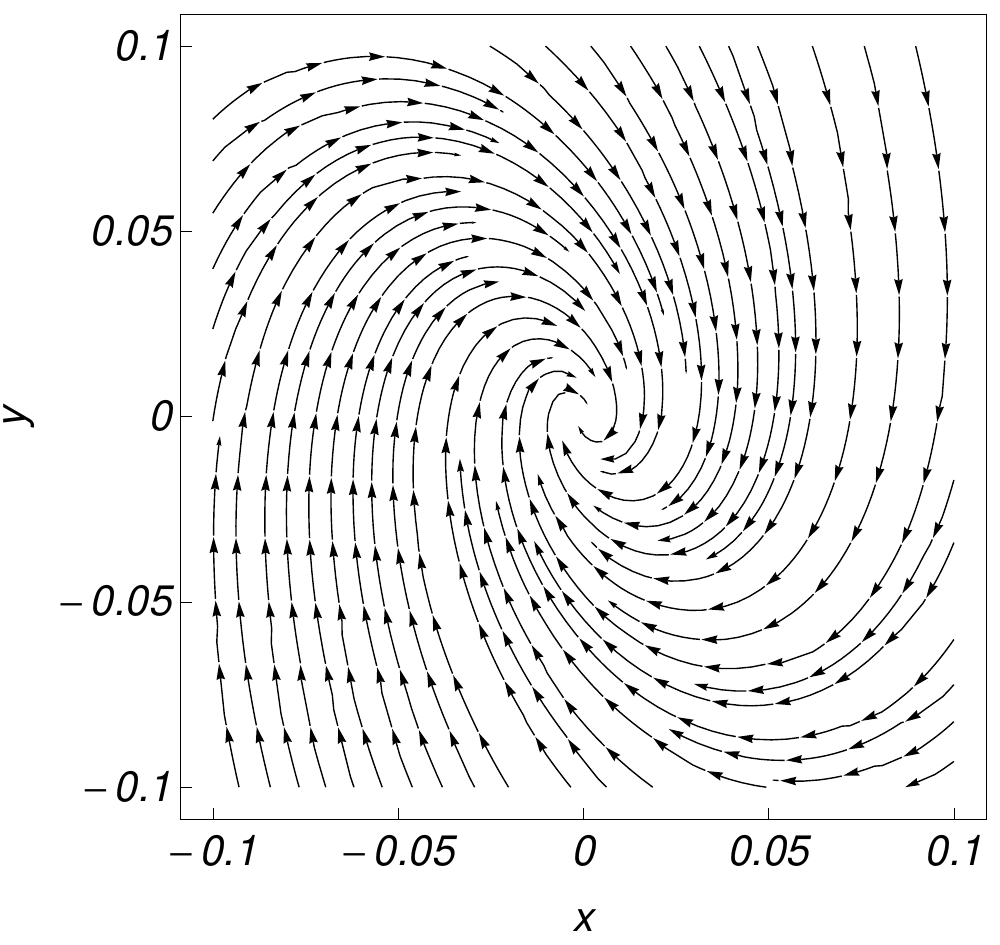}
\caption{Linearized RG flow in the critical plane defined by $r=0$. Arrows point towards the infrared. We set $\vare=1$ for concreteness. \emph{Upper panel.} The couplings $g$ and $\delta$ are attracted to their fixed point values $g_{\star}^2=3\vare/11$ and $\delta_\star=0$, shown here for $x=y=0$. Since the eigenvalue $\theta_\delta=-2\vare/55$ is so small, the coupling $\delta$ is almost constant for long periods of RG time, but eventually vanishes at the infrared fixed point. \emph{Lower panel.} The couplings $x$ and $y$ approach zero in an oscillatory manner, shown here for $g=g_\star$ and $\delta=0$. 
}
\label{FigStream}
\end{minipage}
\end{figure}

On the other hand, for possible realizations of the QBT system with full rotational invariance, such as could be implementations of the Luttinger Hamiltonian (\ref{qbt1}) with ultracold fermions in four distinct hyperfine states, we would have $\delta=0$ and the leading corrections to scaling are given by the complex exponents $\theta_{1,2}$ in Eqs. (\ref{qft20}) and (\ref{qft21}). The finite imaginary parts of these eigenvalues mean that the flow of the couplings is oscillatory in the $x-y$ plane. The linearized flow equations are visualized in Fig. \ref{FigStream}.

\subsection{Dynamical scaling}

Although the term $y \phi^* \partial_\tau\phi$ disappears at the QCP due to $y_\star =0$, the frequency dependence of the boson propagator is still present, being encoded in the coupling $c^2$ that multiplies $\partial_\tau^2$ in Eq. (\ref{qft1}). Note that $c^2$ as a dimensionful coupling is irrelevant at the Gaussian fixed point. From the boson self-energy, however, we find that when the Yukawa coupling is finite, the coupling $c$ is {\it generated} even if absent initially. This can be expressed as  the flow equation
\begin{align}
 \label{rg37} \dot{c^2} = (2-2z-\eta_\phi)c^2  + g^2.
\end{align}
The coefficient $c$ is thus attracted to a small but finite fixed point value
\begin{align}
\label{qft23}  c^2 _\star = \frac{3}{22}\vare.
\end{align}
Hence, whereas $c^2$ vanishes at the Gaussian fixed point, it acquires a nonvanishing value $c>0$ at the interacting QCP. Since the fixed point value of the coefficient $c^2$ is of order of $\vare$, it can be neglected in the lowest order calculation performed here. It would need to be taken into account, however, in the two-loop computation.

The fixed point solution from Eq. (\ref{qft23}) implies that $c$ scales as $c^2\sim \xi^{2z+\eta_\phi-2}$ relative to the diverging correlation length $\xi$. The inverse two-point function (or self-energy) of the bosons
at the transition ($\xi=\infty)$ thus exhibits the scaling form
\begin{align}
 \langle  \phi^* (\omega, \vec p)\, \phi (\omega, \vec p) \rangle^{-1} = p^{2-\eta_\phi} f_\phi\Bigl(\frac{\omega^{1/z}}{p}\Bigr),
\end{align}
where the scaling function $f_\phi$ is such that
\begin{align}
 f_\phi(x) \sim \begin{cases} 1 & x\to 0\\ x^{2-\eta_\phi} & x\to \infty \end{cases}.
\end{align}
Through this mechanism the order parameter inherits the dynamical scaling from the fermions \cite{janssen}.
An analogous scaling form, with $\eta_\phi$ replaced by $\eta_\psi$, can be derived for the two-point function of the fermions.

\section{Discussion}\label{SecDisc}

The field theory studied in this paper bears some resemblance to the theory of the attractive Fermi gas near a Feshbach resonance \cite{nishida,nikolic,radzihovsky,diehl,Zwerger}. In fact, within a particular representation the Yukawa vertex can be written as $\gamma_{45}=\rmi \mathcal{J}$ with the Sp(4) invariant tensor $\mathcal{J}$ from Ref. \cite{nikolic} for a four-component Fermi gas. The main difference, of course, is that the critical point in the atom gas corresponds to zero particle number density, which facilitates an exact computation of the universal quantities. This is not the case here since the Luttinger Hamiltonian is chiral with positive and negative eigenstates, which leads to nonvanishing particle-hole diagrams, such as those in Fig. \ref{FigLoops} a). One therefore has to deal with the full-blown many body problem at criticality, resulting in nontrivial critical exponents.

It may not be too surprising that the cubic anisotropy turned out to be irrelevant at the critical point, as such ``isotropization" is a rather common feature
of RG flow. Indeed, it is difficult to imagine a discrete spatial symmetry alone coexisting with the scale invariance that emerges at the critical point. It is still interesting to observe that the two leading order terms in Eq. (\ref{qft10}) almost cancelled, leaving a rather small eigenvalue behind. In this way, the cubic anisotropy assumes the role of the least irrelevant coupling, which yields the leading correction to scaling, and which is typically reserved for the deviation of the interaction from the fixed point value \cite{herbutbook}. A different scenario has been observed in Ref. \cite{savary}, where the critical point features an emergent anisotropy. As pointed out in the same reference, this is due to the Yukawa vertex being $\gamma_a$ in this case, such that the transpositions appearing in Eq. (\ref{rg7}) and (\ref{rg8}) cannot be removed by commutating through the vertex inside the trace.

Maybe a  more remarkable feature of the field theory in Eq. (\ref{qft7}) is the irrelevance of the particle-hole asymmetry in the infrared. In purely bosonic theories that describe the superfluid-to-Mott insulator transition in the Bose--Hubbard model, for example, the particle-hole symmetry breaking parameter $y$ is relevant at the XY critical point \cite{fisher, herbutbook}. In our case, however, there are two parameters, $x$ and $y$, which together account for the asymmetry,  and which are coupled by the RG flow. Their irrelevance then arises partly as a result of their interplay, and partly due to the fact that the anomalous dimensions are such that  $ \eta_\phi > \eta_\psi$. It is easy to conceive different coefficients in Eqs. (\ref{qft8}) and (\ref{qft9}), hypothetical $\eta_\psi \gg \eta_\phi$ for example, that would yield the opposite result. It is remarkable in this regard that this does not happen. It is also interesting that the particle-hole asymmetry eigenvalues turn out to be complex, which is also, to the best of our knowledge, only rarely the case. (Some examples do exist, but seem to inevitably involve disorder average. See Refs. \cite{weinrib,herbut1998}.) This feature, in particular, also implies that the flow equations cannot be represented as a gradient flow, with an underlying positive-definite metric behind it \cite{wallace}.

\begin{figure}[t]
\centering
\includegraphics[width=4.5cm]{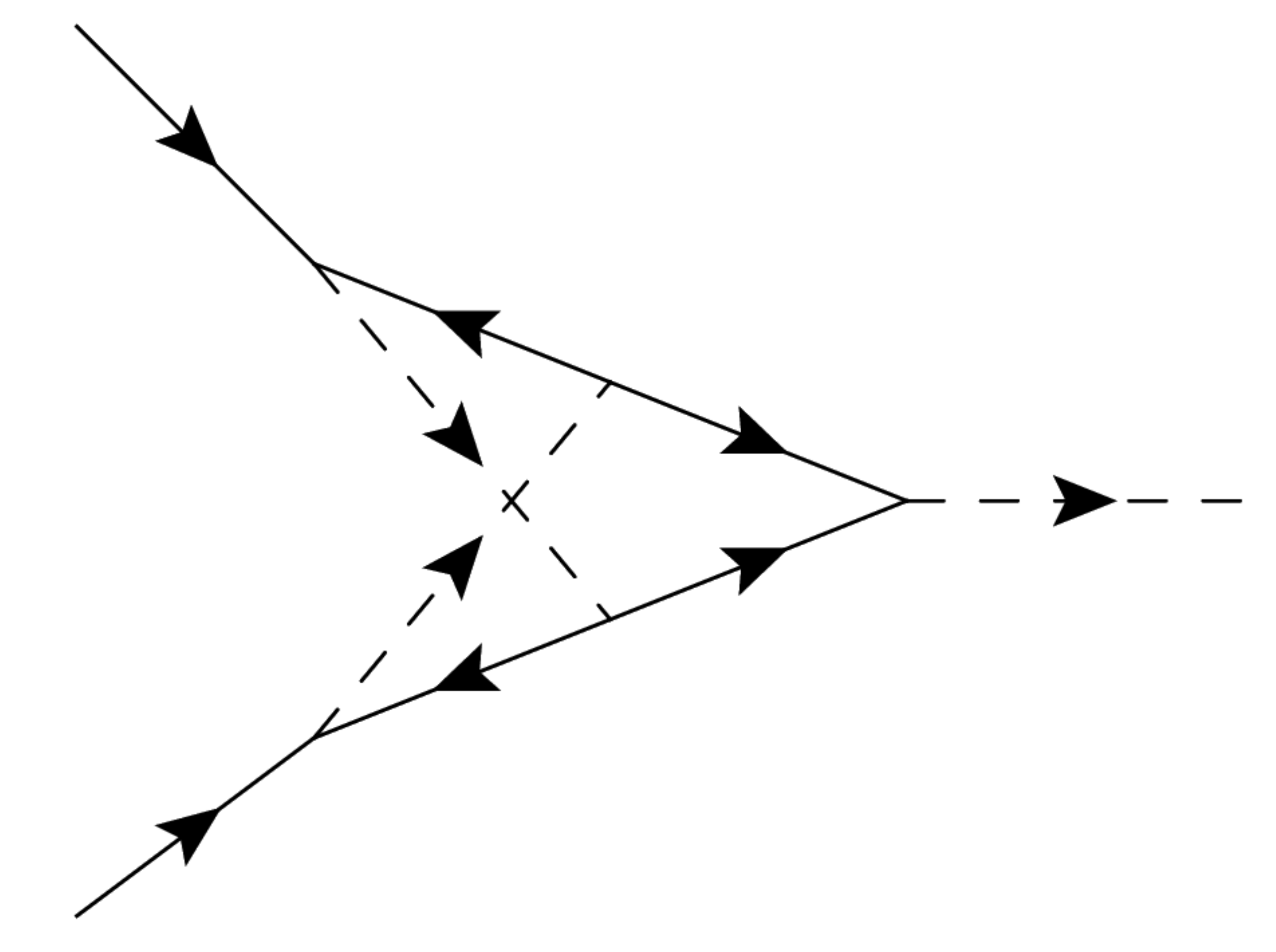}
\caption{The single two-loop diagram that renormalizes the Yukawa interaction vertex directly.}
\label{FigVertex2}
\end{figure}

To the order of our calculation, we found that the sum of the fermionic and bosonic anomalous dimensions is simply $\vare$. More precisely, from Eq. (\ref{qft11}) it follows that
  \begin{equation}
  \eta_\psi + \eta_\phi = \vare + O(\vare^2).
  \end{equation}
We expect the higher order terms, however, to modify this simple looking relation, in contrast to a similar, but exact result in Higgs scalar electrodynamics \cite{tesanovic, litim}.  The two-loop diagram which can be constructed from the Feynman rules of the critical theory and which potentially leads to a direct renormalization of the Yukawa vertex $g$, that is, the finite $O(\vare^2)$-contribution in the above equation, is displayed in Fig. \ref{FigVertex2}.

It is worth noting that quantum fluctuations, as expected in general, in the present case as well {\it suppress}  the superconducting transition temperature. When the chemical potential is right at the QBT, as being the case throughout the paper, the critical temperature near the quantum critical point at $T=0$ and $u=u_{\rm c}$ scales as \cite{herbutbook}
\begin{equation}
T_{\rm c} \sim (u_{\rm c}-u)^{z\nu},
\end{equation}
and we found the relevant combination of the critical exponents to the leading order to be
\begin{equation}
z\nu = \frac{2-\eta_\psi}{2 - \eta_\phi }  = 1 + \frac{7}{22}\vare + O(\vare ^2),
\end{equation}
which is larger than the Gaussian value $z\nu=1$. When the chemical potential is away from the QBT, on the other hand, the critical temperature is finite at an infinitesimal attractive interaction $u$, and assumes the characteristic BCS form
\begin{equation}
\ln T_{\rm c} \propto - \frac{1}{|u| N(\mu)},
\end{equation}
with $N(\mu)$ the density of states at the Fermi level $\mu \neq 0$. At $u=u_{\rm c}$, for example,
\begin{equation}
N (\mu) \sim |\mu | ^{  \frac{d}{z} -1 },
\end{equation}
and is therefore reduced near the QBT by virtue of having $z<2$. Note that, to the leading order, it is the inequality $\eta_\phi > \eta_\psi$ that is related to both the anticipated suppression of $T_{\rm c}$, and to the irrelevance of particle-hole asymmetry (Eq. (\ref{qft9})).

An important ingredient left out in our considerations is the long-range Coulomb repulsion between electrons, which is a relevant perturbation at the Gaussian fixed point of the Luttinger Hamiltonian \cite{abrikosov, moon, herbut2014, janssen2015}.  We have in effect assumed that although the Coulomb interactions cannot be screened when the chemical potential is right at the QBT, we can neglect their growth under RG at the scales of interest due to a sufficiently large dielectric constant of the host crystal. A preliminary calculation suggests, however, that the Coulomb coupling remains relevant at the interacting QCP studied in this paper as well. The full treatment that would include the flow of the Coulomb interaction is somewhat beyond the scope of the present paper, and will be presented in  a future publication.

\begin{center}
 \textbf{Acknowledgements}
\end{center}

\noindent We thank R. Ganesh for inspiring discussions, and particularly Lukas Janssen for a collaboration on an earlier and related project.
I. B. acknowledges funding by the DFG under Grant No. BO 4640/1-1. I. F. H. is supported by NSERC of Canada.

\begin{appendix}

\section{Fierz identities}\label{AppFierz}
For completeness we briefly recall the derivation of the Fierz identities by closely following the presentation given in Ref. \cite{herjurroy}. Let $X$ be the space of Hermitean $4\times 4$ matrices and let $\{\Gamma^A\}_{A=1,\dots,16}$ be an orthogonal basis of $X$ satisfying $\mbox{tr}(\Gamma^A\Gamma^B)=4\delta^{AB}$. Every matrix $M\in X$ can then be uniquely represented as
\begin{align}
 \label{fierz1} M = \frac{1}{4} \mbox{tr}(M\Gamma^A)\Gamma^A.
\end{align}
Written in terms of matrix elements this identity reads $M_{kl} (\Gamma^A)_{lk} (\Gamma^A)_{ij} = 4 M_{ij} = 4 \delta_{ik}\delta_{jl}M_{kl}$. Since $M$ was arbitrary we conclude that the relation
\begin{align}
 \label{fierz2} \delta_{ik}\delta_{jl}= \frac{1}{4}(\Gamma^A)_{ij} (\Gamma^A)_{lk}
\end{align}
holds between the basis vectors $\Gamma^A$. Given two matrices $M, N\in X$ we consider the expression
\begin{align}
 \label{fierz3} \mbox{tr}(M\Gamma^A N \Gamma^B) = M_{ij} N_{kl} (\Gamma^A)_{jk}(\Gamma^B)_{li}.
\end{align}
Contracting both sides with $(\Gamma^A)_{mn}(\Gamma^B)_{op}$ and using Eq. (\ref{fierz2}) we arrive at
\begin{align}
 \label{fierz4} M_{ij}N_{mn} = \frac{1}{16}\mbox{tr}(M\Gamma^AN\Gamma^B)(\Gamma^A)_{mj}(\Gamma^B)_{in}.
\end{align}
Fierz identities consist in the application of Eq. (\ref{fierz4}) to products of fermion bilinears. Given $M,N\in X$ and $\psi_x:=\psi(\tau,\vec{x})=(\psi_1,\psi_2,\psi_3,\psi_4)^T$ one readily finds
\begin{align}
 \nonumber (\psi_x^\dagger M \psi_x)(\psi_y^\dagger N \psi_y)  ={}& -\frac{1}{16}\mbox{tr}(M\Gamma^AN\Gamma^B)\\
 \label{fierz5} &\times  (\psi_y^\dagger\Gamma^A\psi_x)(\psi_x^\dagger\Gamma^B\psi_y),\\
 \nonumber (\psi^\dagger_x M \psi^*_x)(\psi^{\rm T}_y N \psi_y) ={}&  \frac{1}{16}\mbox{tr}(M\Gamma^AN\Gamma^B) \\
 \label{fierz6} &\times  (\psi^\dagger_x (\Gamma^A)^{\rm T} \psi_y) (\psi^\dagger_x \Gamma^B \psi_y).
\end{align}

We now aim at relating the term $(\psi^\dagger\psi)^2$ to $L_{\rm s}, L_{\rm d}$ in Eqs. (\ref{qbt7}), (\ref{qbt8}) by means of the Fierz identity at $\vec{x}=\vec{y}$. For this purpose we choose
\begin{align}
 \label{fierz7} \{\Gamma^A\} = \Bigl\{ \mathbb{1},\ \{\gamma_a\}_{a=1,\dots,5},\ \{\gamma_{ab}=\rmi \gamma_a\gamma_b\}_{a<b}\Bigr\},
\end{align}
which satisfies $\mbox{tr}(\Gamma^A\Gamma^B)=4\delta^{AB}$. For the s-wave term $L_{\rm s}$ we have $M=N=\gamma_{45}$ in Eq. (\ref{fierz6}),  and find
\begin{align}
 \label{fierz8} \mbox{tr}(\gamma_{45}\Gamma^A\gamma_{45}\Gamma^B) = 4\sigma_A \delta^{AB},
\end{align}
where $\sigma_A$ is such that $(\Gamma^A)^{\rm T}=\sigma_A\Gamma^A$ for $\Gamma^A\in\{\mathbb{1},\gamma_a\}$, whereas $(\Gamma^A)^{\rm T}=-\sigma_A\Gamma^A$ for $\Gamma^A\in\{\gamma_{ab}\}$. Consequently,
\begin{align}
 \nonumber L_{\rm s} &= \frac{1}{4} \sum_A \sigma_A (\psi^\dagger (\Gamma^A)^{\rm T}\psi)(\psi^\dagger \Gamma^A \psi)\\
 \label{fierz9} &=\frac{1}{4}\Bigl[(\psi^\dagger\psi)^2 +(\psi^\dagger\gamma_a\psi)^2-(\psi^\dagger\gamma_{ab}\psi)^2\Bigr].
\end{align}
Note that when writing $(\psi^\dagger\gamma_{ab}\psi)^2$ we implicitly assume a sum over $a<b$ only. In the case of the d-wave term $L_{\rm d}$ we have $M=\gamma_a\gamma_{45}$ and $N=M^\dagger=\gamma_{45}\gamma_a$. Analogous to the s-wave case we find
\begin{align}
 \label{fierz12} L_{\rm d} &=\frac{1}{4}\Bigl[5(\psi^\dagger\psi)^2 -3(\psi^\dagger\gamma_a\psi)^2-(\psi^\dagger\gamma_{ab}\psi)^2\Bigr].
\end{align}

With Eqs. (\ref{fierz9}) and (\ref{fierz12}) we have rewritten the superconducting terms $L_{\rm s}$ and $L_{\rm d}$ as a linear combination of the bilinears $(\psi^\dagger\psi)^2$, $(\psi^\dagger\gamma_a\psi)^2$, and $(\psi^\dagger\gamma_{ab}\psi)^2$. One of these terms can be eliminated, however, by expressing it in terms of the other two.
For this we evaluate Eq. (\ref{fierz5}) for $M=N=\mathbb{1}$ to obtain
\begin{align}
 \label{fierz13} (\psi^\dagger\psi)^2 &= -\frac{1}{4}(\psi^\dagger\Gamma^A\psi)^2,
\end{align}
hence
\begin{align}
 \label{fierz16} L_{\rm s}+L_{\rm d} = \frac{1}{4}\Bigl[8(\psi^\dagger\psi)^2-2(\psi^\dagger\Gamma^A\psi)^2\Bigr]=4(\psi^\dagger\psi),
\end{align}
as stated in Eq. (\ref{qbt6}) in the main text. One can show that no further reduction of the fermion bilinears is possible \cite{herbut2014}.

\section{Renormalization group equations}\label{AppB}

In this appendix we provide details on the derivation of the RG equations (\ref{qft8})-(\ref{qft14}) which are used in the main text to study the superconducting QCP. We first recall general facts about the anomalous scaling of running couplings \cite{herbutbook}, and then derive the RG equations for the cases of $\delta=0$ and $\delta \neq0$, separately. Within this section $N=4$ is the number of fermion components.

\subsection{Anomalous scaling}\label{AppScal}

To discuss the scaling properties of running couplings we consider the Lagrangian close to the transition given by
\begin{align}
 \nonumber  \bar{L}(\bar{\psi},\bar{\phi}) = {}& \bar{\psi}^\dagger(\bar{S}\partial_{\bar{\tau}}+A_\psi d_a(-\rmi \nabla)\gamma_a-\bar{x}\nabla^2)\bar{\psi} \\
 &\nonumber +\bar{\phi}^*(\bar{y}\partial_{\bar{\tau}}-A_\phi \nabla^2-\bar{c}^2\partial_{\bar{\tau}}^2+\bar{r})\bar{\phi}\\
 \label{rg1} &+ \bar{g} ( \bar{\phi} \bar{\psi}^\dagger \gamma_{45} \bar{\psi}^* + \bar{\phi}^* \bar{\psi}^{\rm T} \gamma_{45} \bar{\psi}).
\end{align}
For simplicity we assume $\delta=0$. Notice that we have not set $S=A_\psi=A_\phi=1$ in Eq. (\ref{rg1}) and label this choice of parameters by an overbar to the quantities.

Since we are free to re-parametrize the fields $\bar{\phi}$ and $\bar{\psi}$ in $\bar{L}$ we first choose them such that the prefactors of $\psi^\dagger d_a\gamma_a \psi$ and $\phi^*(-\nabla^2)\phi$ remain unity during the RG procedure. This is accomplished by choosing $\hat{\psi}=A_\psi^{1/2}\bar{\psi}$ and $\hat{\phi}=A_\phi^{1/2}\bar{\phi}$. We then find
\begin{align}
 \nonumber  \bar{L}_{\rm kin}(\hat{\psi},\hat{\phi}) = {}& \hat{\psi}^\dagger\Bigl(\frac{\bar{S}}{A_\psi} \partial_{\bar{\tau}}+ d_a(-\rmi \nabla)\gamma_a-\frac{\bar{x}}{A_\psi}\nabla^2\Bigr)\hat{\psi} \\
 \label{rg1b}&+\hat{\phi}^*\Bigl(\frac{\bar{y}}{A_\phi}\partial_{\bar{\tau}}- \nabla^2-\frac{\bar{c}^2}{A_\phi}\partial_{\bar{\tau}}^2\Bigr)\hat{\phi}.
\end{align}
for the kinetic term. Let us denote $S=\bar{S}/A_\psi$. Due to the absence of Lorentz invariance in the non-relativistic setting we have the additional freedom to choose an adjusted time-coordinate $\tau=\bar{\tau}/S$ such that the prefactor of $\psi^\dagger \partial_\tau \psi$ is kept at unity as well. The kinetic part of the action, $\bar{\mathcal{S}}_{\rm kin}= \int \mbox{d}\bar{\tau}\mbox{d}^dx\ \bar{L}_{\rm kin}$, remains invariant under this transformation if we rescale fields a second time according to $\psi=S ^{1/2}\hat{\psi}$ and $\phi=S^{1/2}\hat{\phi}$. The interaction part of the action becomes $\int\mbox{d}\tau \mbox{d}^dx\ \bar{g} S A_\phi^{-1/2}A_\psi^{-1}S^{-3/2}(\phi \psi^\dagger\gamma_{45}\psi^*+\text{h.c.})$. The Lagrangian in the new field coordinates is given by
\begin{align}
 \nonumber  L(\psi,\phi) = {}& \psi^\dagger\Bigl(\partial_{\tau}+ d_a(-\rmi \nabla)\gamma_a-\frac{\bar{x}}{A_\psi}\nabla^2\Bigr)\psi \\
 &\nonumber +\phi^*\Bigl(\frac{\bar{y}}{A_\phi S}\partial_{\tau}- \nabla^2-\frac{\bar{c}^2}{A_\phi S^2}\partial_{\tau}^2+\frac{\bar{r}}{A_\phi}\Bigr)\phi\\
 \label{rg1c} &+ \frac{\bar{g}}{A_\psi A_\phi^{1/2}S^{1/2}} ( \phi \psi^\dagger \gamma_{45} \psi^* + \phi^* \psi^{\rm T} \gamma_{45} \psi).
\end{align}
We conclude that the Lagrangian can be brought into the form (\ref{qft1}) by proper re-parametrization of the fields, and that the couplings of the rescaled theory acquire multiplicative factors of $A_\psi$, $A_\phi$, and $S$. This manifests in additive anomalous scaling terms in the beta functions, as we outline here.

Consider an arbitrary coupling $\bar{\lambda}$ (such as $\bar{x}$, $\bar{y}$, $\bar{g}$, $\dots$) with scaling dimension $d_1=\text{dim}[\bar{\lambda}]$. The flow equation for $\bar{\lambda}$ will then have the form
\begin{align}
 \label{rg2} \dot{\bar{\lambda}} = d_1 \bar{\lambda} + \beta_{\bar{\lambda}}.
\end{align}
Let us denote the rescaled coupling schematically by $\lambda=\bar{\lambda} A_\psi^{-d_2}A_\phi^{-d_3}S^{-d_4}$. The flow equation for $\lambda$ is then given by
\begin{align}
 \nonumber \dot{\lambda} &= b \frac{\mbox{d}}{\mbox{d}b} \Bigl(\frac{\bar{\lambda}}{A_\psi^{d_2}A_\phi^{d_3}S^{d_4}}\Bigr)\\
 \nonumber &= \Bigl(d_1 -d_2 \frac{\dot{A}_\psi}{A_\psi}-d_3\frac{\dot{A}_\phi}{A_\phi} -d_4\frac{\dot{S}}{S}\Bigr)\lambda +\frac{\beta_{\bar{\lambda}}}{A_\psi^{d_2}A_\phi^{d_3}S^{d_4}}\\
 \label{rg3} &=:\Bigl(d_1 -d_2 \eta_\psi-d_3\eta_\phi -d_4(z-2)\Bigr)\lambda  +\beta_\lambda,
\end{align}
where we defined the fermion and boson anomalous dimensions by
\begin{align}
  \label{rg4} \eta_\psi = \frac{1}{A_\psi} \dot{A}_\psi,\ \eta_\phi = \frac{1}{A_\phi} \dot{A}_\phi,
\end{align}
and the dynamical critical exponent $z$ by means of
\begin{align}
 \label{rg5} \dot{S} = (z-2)S.
\end{align}
The appearance of $\eta_\phi$, $\eta_\psi$ and $\eta_S=z-2$ on the same footing as the canonical dimension $d_1$ justifies calling them anomalous dimensions. By rescaling all running couplings of the theory by appropriate powers of $A_\psi$, $A_\phi$, and $S$, the latter three will eventually disappear from the beta functions. For practical purposes it is convenient to compute $\beta_{\bar{\lambda}}$ by setting $A_\psi=A_\phi=1$ from the outset and add the anomalous scaling terms by hand. We apply this procedure in the following. However, we keep track of $S\neq 1$ and trade it for $z$ at the end of the calculation.

From the construction just presented it is clear that we might equally well choose the time-coordinate such that the prefactor of $\phi^* \partial_\tau^2\phi$ remains unity, and the corresponding dynamical critical exponent of bosons thus necessarily coincides with the fermionic one \cite{janssen2015}.

\subsection{RG flow with full rotation invariance}\label{AppRG}

We now derive the RG equations assuming  $\delta=0$. By integrating out fluctuations in the momentum shell $[\Lambda/b,\Lambda]$ to one-loop order we generate a contribution to the effective Lagrangian (\ref{rg1}) given by
\begin{align}
 \label{rg6} \delta \bar{L} = \bar{\psi}^\dagger \Sigma_\psi \bar{\psi} + \bar{\phi}^* \Sigma_\phi \bar{\phi}.
\end{align}
The additional terms $\Sigma_\psi$ and $\Sigma_\phi$ constitute loop corrections to the fermion and boson self-energies, respectively. Their graphic representation is shown in Fig. \ref{FigLoops}. Note that there is no renormalization of the coupling $\bar{g}$ to one-loop order as no such diagram can be constructed from the Feynman rules dictated by $L$. The explicit expressions for the self-energy corrections are
\begin{align}
 \label{rg7} \Sigma_\psi(P) &= 4\bar{g}^2 \int_Q \frac{1}{\mathcal{P}_\phi^{Q+P}\mbox{det}_M^Q} \Bigl(\gamma_{45}[\mathcal{M}_\psi^{-Q}]^{\rm T}\gamma_{45}\Bigr),\\
 \label{rg8}\Sigma_\phi(P) &= -2\bar{g}^2 \int_Q \frac{\mbox{tr}(\gamma_{45}\mathcal{M}_\psi^{-Q}\gamma_{45}[\mathcal{M}_\psi^{Q-P}]^{\rm T})}{\mbox{det}_M^Q\mbox{det}_M^{P-Q}},
\end{align}
where $P=(p_0,\vec{p})$ and we abbreviate
\begin{align}
 \label{rg9} \int_Q = \int_{-\infty}^\infty \frac{\mbox{d}q_0}{2\pi} \int_{\vec{q}}^\prime,\ \int_{\vec{q}}^\prime = \int_{\Lambda/b}^\Lambda \frac{\mbox{d}^dq}{(2\pi)^d}.
\end{align}
The prime indicates that the $\vec{q}$-integration is limited to the momentum shell $\Lambda/b\leq q\leq \Lambda$.  In Eqs. (\ref{rg7}) and (\ref{rg8}) we have
\begin{align}
 \label{rg10} \mathcal{P}_\phi^Q &= \rmi \bar{y} q_0 +q^2,\\
 \label{rg11} \mathcal{M}_\psi^Q &=(S\rmi q_0-\bar{x}q^2)\mathbb{1}_N +d_a(\vec{q})\gamma_a,\\
 \label{rg12} \mbox{det}_M^Q &= S^2 q_0^2 +(1-\bar{x}^2)q^4 -2 (S\rmi q_0)(\bar{x}q^2).
\end{align}
We generalized to $N$ fermion components, where $N$ is a multiple of four.

The expressions for the self-energy corrections can be simplified by commutating through the vertices $\gamma_{45}$. Indeed, in our representation where $\gamma_{1,2,3}$ are real and $\gamma_{4,5}$ are imaginary we have
\begin{align}
 \label{rg13} (\gamma_a)^{\rm T} = [-1]^a\gamma_a,\ [-1]^a =\begin{cases} 1 & a=1,2,3\\ -1 & a=4,5\end{cases},
\end{align}
which follows from $\gamma_a^\dagger=\gamma_a$. At the same time we have
\begin{align}
 \label{rg14} \gamma_{45}\gamma_a = [-1]^a \gamma_a \gamma_{45},
\end{align}
following from the fact that $\gamma_{45}$ is proportional to the product of $\gamma_4$ and $\gamma_5$. Thus we have
\begin{align}
 \label{rg15} \gamma_{45} [\mathcal{M}_\psi^Q]^{\rm T}\gamma_{45} = \mathcal{M}_\psi^Q (\gamma_{45})^2 = \mathcal{M}_\psi^Q.
\end{align}
We are left with the expressions
\begin{align}
 \label{rg16} \Sigma_\psi(P) &= 4\bar{g}^2 \int_Q \frac{1}{\mathcal{P}_\phi^{Q+P}\mbox{det}_M^Q}\mathcal{M}_\psi^{-Q},\\
 \label{rg17}\Sigma_\phi(P) &= -2\bar{g}^2\int_Q \frac{1}{\mbox{det}_M^Q\mbox{det}_M^{P-Q}}\mbox{tr}(\mathcal{M}_\psi^{-Q}\mathcal{M}_\psi^{Q-P}).
\end{align}
Performing the trace in $\Sigma_\phi$ yields
\begin{align}
 \nonumber  \mbox{tr}\Bigl(\mathcal{M}_\psi^{-Q}\mathcal{M}_\psi^{Q-P}\Bigr) = N \Bigl[{}& (-\rmi S q_0-\bar{x} q^2)\Bigl(\rmi S (q_0-p_0)\\
 \label{rg18} &-\bar{x}(\vec{q}-\vec{p})^2\Bigr)+d_a(\vec{q})d_a(\vec{q}-\vec{p})\Bigr].
\end{align}

We now employ an expansion of Eqs. (\ref{rg16}) and (\ref{rg17}) in powers of $\rmi p_0$ and $p^2$ to read off the one-loop contribution to the couplings of $L_{\rm c}$. In the loops, the frequency integration can be evaluated analytically. The momentum integration is facilitated by means of the $d$-function technology presented in App. \ref{AppdTech} and
\begin{align}
 \label{rg19} \int_{\vec{q}}^\prime \frac{1}{q^4}  = \frac{\text{S}_d}{(2\pi)^d} \Lambda^{d-4}\log b,
\end{align}
which is valid for all $b$ as $\vare\to0$. Herein, $\text{S}_d$ is the surface area of the $d$-dimensional unit ball. Furthermore, on the right hand side of the flow equations we set $r=0$, thus assuming the theory to be close to its critical point, and neglect the feedback of $\bar{c}^2$, which is an irrelevant coupling.

To explicate this procedure we first compute the contributions to the beta functions which result from $\Sigma_\psi$. They are technically simple as the external momenta and frequencies can be shifted to the boson propagator. We expand the latter to order $p_0$ and $p^2$ according to
\begin{align}
 \label{rg19b} \frac{1}{\mathcal{P}_\phi^{Q+P}} \simeq \frac{1}{\mathcal{P}_\phi^{Q}}\Bigl(1-\frac{\rmi \bar{y}p_0+2\vec{q}\cdot\vec{p}+p^2}{\mathcal{P}_\phi^Q}+\frac{4(\vec{q}\cdot\vec{p})^2}{(\mathcal{P}_\phi^Q)^2}\Bigr),
\end{align}
and arrive at
\begin{align}
 \nonumber  \Sigma_\psi(P) \simeq \Sigma_\psi(0) &+ \frac{2S(\rmi \bar{y}p_0+p^2)\bar{g}^2}{[S+(1-\bar{x})\bar{y}]^2} \mathbb{1}_N \int_{\vec{q}}^\prime\frac{1}{q^4}\\
 \nonumber &- \frac{8S^2\bar{g}^2}{[S+(1-\bar{x})\bar{y}]^3} \mathbb{1}_N \int_{\vec{q}}^\prime \frac{(\vec{q}\cdot\vec{p})^2}{q^6}\\
 \nonumber &- \frac{2S(\rmi \bar{y}p_0+p^2)\bar{g}^2}{[S+(1-\bar{x})\bar{y}]^2} \gamma_a \int_{\vec{q}}^\prime \frac{\mbox{d}_a(\vec{q})}{q^6}\\
 &+ \frac{8S^2\bar{g}^2}{[S+(1-\bar{x})\bar{y}]^3}\gamma_a \int_{\vec{q}}^\prime \frac{(\vec{q}\cdot\vec{p})^2d_a(\vec{q})}{q ^8}.
\end{align}
We have eliminated terms that are odd in the components $q_i$ as the integration domain $\Lambda/b\leq q \leq \Lambda$ is even under $q_i\to-q_i$ for fixed $i$. In order to compute the last line we make use of Eq. (\ref{d4}) with $d=4$ and write
\begin{align}
 \label{rg19e} (\vec{q}\cdot\vec{p})^2  = \frac{3}{4} d_b(\vec{q})d_b(\vec{p})+\frac{1}{4}q^2p^2.
\end{align}
The remaining integrals can then be simplified by using the results of App. \ref{AppdTech} for $d=4$. Here we employ
\begin{align}
 \label{rg19dd} \int_{\vec{q}}f(q^2) (\vec{q}\cdot\vec{p})^2 & =\frac{1}{4} p^2 \int_{\vec{q}} f(q^2) q^2,\\
 \label{rg19de} \int_{\vec{q}}f(q^2) d_a(\vec{q}) &= 0,\\
 \label{rg19df} \int_{\vec{q}}f(q^2) d_a(\vec{q})d_b(\vec{q}) &= \frac{1}{9} \delta_{ab} \int_{\vec{q}}f(q^2) q^4,
\end{align}
where $f(q^2)$ is some function that has compact support on the momentum shell $\Lambda/b\leq q \leq \Lambda$, and arrive at
\begin{align}
 \nonumber \Sigma_\psi(P) = \Sigma_\psi(0) &+\frac{2S(\rmi \bar{y}p_0+p^2)\bar{g}^2}{[S+(1-\bar{x})\bar{y}]^2}\mathbb{1}_N\int_{\vec{q}}^\prime \frac{1}{q^4}\\
 \nonumber &-\frac{2S^2p^2\bar{g}^2}{[S+(1-\bar{x})\bar{y}]^3}  \mathbb{1}_N \int_{\vec{q}}^\prime \frac{1}{q^4},\\
 \label{rg19g} &+ \frac{2S^2\bar{g}^2}{3[S+(1-\bar{x})\bar{y}]^3} d_a(\vec{p})\gamma_a \int_{\vec{q}}^\prime \frac{1}{q^4}.
\end{align}
After employing Eq. (\ref{rg19}), the loop corrections to $\bar{S}$, $\bar{x}$, and $A_\psi$ can now be read off as the coefficients multiplying $\rmi p_0 \mathbb{1}_N$, $p^2 \mathbb{1}_N$, and $d_a(\vec{p})\gamma_a$, respectively.

To compute the boson self-energy $\Sigma_\phi(P)$ in Eq. (\ref{rg17}) requires expanding $\mathcal{M}_\psi^{Q-P}$ and $\mbox{det}_M^{P-Q}$ in powers of $\rmi p_0$ and $p^2$. In Eq. (\ref{rg18}) we employ $d_a(\vec{q})d_a(\vec{q}-\vec{p})=\frac{1}{3}\{4[\vec{q}\cdot(\vec{q}-\vec{p})]^2-q^2(\vec{q}-\vec{p})^2\}$. We then find
\begin{align}
 \label{rg19h} \frac{\partial \Sigma_\phi}{\partial \rmi p_0}\Bigr|_{P=0} = -\frac{N\bar{g}^2\bar{x}}{(1-\bar{x}^2)^2}\int_{\vec{q}}^\prime \frac{1}{q^4}.
\end{align}
In the same way, the contribution to $\bar{c}^2$ can easily be derived.  To extract the $p^2$-dependence we set $p_0=0$, replace $\vec{p}\to t \vec{p}$, and compute
\begin{align}
 \label{rg19i} I_t=\frac{1}{2}\frac{\partial^2\Sigma_\phi}{\partial t^2}\Bigr|_{p_0=t=0} = \frac{N\bar{g}^2}{6S(1-\bar{x}^2)}\int_{\vec{q}}^\prime \frac{7p^2q^2-10(\vec{q}\cdot\vec{p})^2}{q^6}.
\end{align}
We make again use of Eq. (\ref{rg19dd}) to arrive at
\begin{align}
 \label{rg19j} \frac{\partial \Sigma_\phi}{\partial p^2}\Bigr|_{P=0} = \frac{\partial I_t}{\partial p^2} = \frac{3N\bar{g}^2}{4S(1-\bar{x}^2)}\int_{\vec{q}}^\prime \frac{1}{q^4}.
\end{align}
The method of introducing the parameter $t$ will be convenient for the evaluation of $\Sigma_\phi(P)$ with $\delta\neq 0$ in App. \ref{AppAniso}.

With these findings we eventually arrive at the one-loop corrections
\begin{align}
 \label{rg20} \bar{S}(b) &= 1 +\frac{2S\bar{y}}{[S+(1-\bar{x})\bar{y}]^2} \frac{\text{S}_d\Lambda^{d-4}\bar{g}^2}{(2\pi)^d}\log b,\\
 \label{rg21} A_\psi(b) &= 1+\frac{2S^2}{3[S+(1-\bar{x})\bar{y}]^3}\frac{\text{S}_d\Lambda^{d-4}\bar{g}^2}{(2\pi)^d}\log b,\\
 \label{rg22} \bar{x}(b) &= \bar{x}(1) + \frac{2S(1-\bar{x})\bar{y}}{[S+(1-\bar{x})y]^3} \frac{\text{S}_d\Lambda^{d-4}\bar{g}^2}{(2\pi)^d} \log b,\\
 \label{rg23}  A_\phi(b) &= 1+ \frac{3N}{4S(1-\bar{x}^2)}\frac{\text{S}_d\Lambda^{d-4}\bar{g}^2}{(2\pi)^d}\log b,\\
 \label{rg24} \bar{y}(b) &= \bar{y}(1) -\frac{N \bar{x}}{(1-\bar{x}^2)^2} \frac{\text{S}_d\Lambda^{d-4}\bar{g}^2}{(2\pi)^d} \log b,\\
 \label{rg25} \bar{g}^2(b) &= b^{4-d} \bar{g}^2(1),\\
 \label{rg25b} \bar{r}(b) &= b^2\Bigl(\bar{r}(1)-\frac{N\Lambda^2}{2S(1-\bar{x}^2)} \frac{\text{S}_d\Lambda^{d-4}\bar{g}^2}{(2\pi)^d}(1-b^{2-d})\Bigr),\\
 \label{rg26} \bar{c}^2(b) &= \frac{1}{b^2}\Bigl(\bar{c}^2(1)-\frac{NS(1+3\bar{x}^2)}{8\Lambda^2(1-\bar{x}^2)^3} \frac{\text{S}_d \Lambda^{d-4}\bar{g}^2}{(2\pi)^d}(1-b^{6-d})\Bigr).
\end{align}
We introduce the rescaled couplings
\begin{align}
 \label{newcouplings}  S&=\frac{\bar{S}}{A_\psi}=1,\ x = \frac{\bar{x}}{A_\psi},\ y=\frac{\bar{y}}{A_\phi S},\ g^2 = \frac{\text{S}_d\Lambda^{d-4}}{(2\pi)^d} \frac{\bar{g}^2}{A_\psi^2A_\phi S}
\end{align}
to arrive at the RG equations
\begin{align}
 \label{rg27} \dot{x} &= -\eta_\psi x+ \frac{2(1-x)yg^2}{[1+(1-x)y]^3},\\
 \label{rg28} \dot{y} &= (2-z-\eta_\phi) y -\frac{N x g^2}{(1-x^2)^2},\\
 \label{rg29} \dot{g} &= \frac{1}{2}(2+\vare -\eta_\phi-2\eta_\psi-z)g,
\end{align}
with
\begin{align}
 \label{rg30} \eta_\phi &= \frac{3N g^2}{4(1-x^2)},\ \eta_\psi = \frac{2g^2}{3[1+(1-x)y]^3},\\
 \label{rg31} z &= 2-\eta_\psi +\frac{2yg^2}{[1+(1-x)y]^2}.
\end{align}
Note that $x< 1$. By expanding these expressions for small $x$ and $y$ we obtain Eqs. (\ref{qft8}), (\ref{qft9}), (\ref{qft10})-(\ref{qft14}) employed in the fixed point analysis.

\subsection{Cubic lattice symmetry}\label{AppAniso}
Next, we consider the influence of a finite anisotropy parameter $\delta>0$ onto the running couplings of
\begin{align}
 \nonumber L ={}& \psi^\dagger\Bigl(\partial_\tau+\sum_a d_a\gamma_a + \delta\sum_a s_ad_a\gamma_a-x \nabla^2\Bigr)\psi\\
 \label{delta1} &+\phi^*(y\partial_\tau-\nabla^2)\phi+ g(\phi\psi^\dagger\gamma_{45}\psi^*+\phi^*\psi^{\rm T}\gamma_{45}\psi).
\end{align}
We recall the definition of $s_a$ as
\begin{align}
 \label{delta7} s_a =\begin{cases} +1 & \Lambda^a\ \text{is off-diagonal},\\ -1 & \Lambda^a\ \text{is diagonal}.\end{cases}
\end{align}
Together with $x$ and $y$, the coupling $\delta$ constitutes a marginal coupling which could be potentially relevant at the QCP with $\delta_\star= x_\star=y_\star=0$. In this section we derive the corrections due to $\delta\neq 0$ onto $\Sigma_\psi $ and $\Sigma_\phi$. We find that $\delta$ is irrelevant and attracted to the fixed point value $\delta_\star=0$ at the QCP. Furthermore, the corrections to $\dot{x}$ and $\dot{y}$ are of order $O(y\delta)$ and $O(x\delta)$, respectively. Hence, the derivation of flow equations from App. \ref{AppRG} where we set $\delta=0$ remains valid for the study of the fixed point structure of the QCP, which only requires the flow equations for small $x, y, \delta$. Note that, in the spirit of the discussion in the beginning of App. \ref{AppScal}, we may first introduce the coupling $\bar{\delta}$ in $\bar{L}$ and then define $\delta=\bar{\delta}/A_\psi$.

Below we show that the corrections to $\Sigma_\psi$ and $\Sigma_\phi$ for small $x,y$ to order $O(\delta^2)$ are given by
\begin{align}
 \label{delta2} \Delta\Sigma_\psi(P) &=  \frac{8}{15}g^2 \delta\ \Bigl[-\frac{1}{3}\sum_ad_a\gamma_a +  \sum_a s_ad_a\gamma_a\Bigr]\log b,\\
 \label{delta3} \Delta \Sigma_\phi(P) &= - g^2\delta\ p^2 \log b,
\end{align}
where we subtracted the $\delta$-independent part according to
\begin{align}
\Delta \mathcal{O} :=\mathcal{O} -\mathcal{O}_{\delta=0}.
\end{align}
This implies the corrections
\begin{align}
 \label{delta4} \Delta \eta_\psi = -\frac{8}{45}g^2\delta,\ \Delta\eta_\phi=-g^2\delta.
\end{align}
Furthermore, the flow equation for $\delta$ is given by
\begin{align}
 \label{delta5} \dot{\delta} = -\eta_\psi \delta + \frac{8}{15} g^2\delta.
\end{align}
The latter equation yields a fixed point of $\delta$ at $\delta_\star=0$. Close to the fixed point the linearized flow reads
\begin{align}
 \label{delta6} \dot{\delta} \approx \Bigl(-\eta_{\psi\star}+\frac{8}{15}g^2_\star\Bigr) \delta = \Bigl(-\frac{2}{3}+\frac{8}{15}\Bigr)g^2_\star \delta = -\frac{2}{55}\vare\ \delta.
\end{align}
Thus, $\delta$ is an irrelevant coupling with scaling dimension $\theta_\delta=-(2/55)\vare$. The implications of this finding are discussed in the main text.

We now compute the self-energy corrections for nonzero $\delta$. In this situation, Eqs. (\ref{rg16}), (\ref{rg17}) are modified according to
\begin{align}
 \label{delta27} \Sigma_\psi(P) &= 4\bar{g}^2 \int_Q \frac{1}{\mathcal{P}_\phi^{Q+P}\mbox{det}_{M,\delta}^Q}\mathcal{M}_{\psi,\delta}^{-Q},\\
 \label{delta28}\Sigma_\phi(P) &= -2\bar{g}^2 \int_Q \frac{1}{\mbox{det}_{M,\delta}^Q\mbox{det}_{M,\delta}^{P-Q}}\mbox{tr}(\mathcal{M}_{\psi,\delta}^{-Q}\mathcal{M}_{\psi,\delta}^{Q-P})
\end{align}
with
\begin{align}
 \label{delta29} \mathcal{M}_{\psi,\delta}^Q &=\mathcal{M}_{\psi}^Q+\delta \sum_a s_a d_a(\vec{q})\gamma_a,\\
 \nonumber \mbox{det}_{M,\delta}^Q &= \mbox{det}_M^Q +\delta \sum_a s_a (2+\delta s_a) d_a^2(\vec{q})\\
 \label{delta30} &\simeq\mbox{det}_M^Q +2\delta \sum_a s_a d_a^2(\vec{q}).
\end{align}
In the last line we have expanded the expression to linear order in $\delta$. This expansion to order $O(\delta^2)$ is also applied in the remainder of this section. From the self-energy expressions (\ref{delta27}) and (\ref{delta28}) we deduce
\begin{align}
 \label{delta31} \Delta \Sigma_\psi(P) &\simeq 4\bar{g}^2 \delta \int_Q \frac{1}{\mathcal{P}_\phi^{Q+P}\mbox{det}_M^Q}\Bigl(\sum_a s_a d_a(\vec{q})\gamma_a\Bigr)\\
 \nonumber &-8\bar{g}^2\delta \int_Q \frac{1}{\mathcal{P}_\phi^{Q+P}(\mbox{det}_M^Q)^2}\Bigl(\sum_a s_a d^2_a(\vec{q})\Bigr)\mathcal{M}_\psi^{-Q}
\end{align}
and
\begin{align}
 \label{delta32} \Delta \Sigma_\phi(P) &\simeq -4N \bar{g}^2 \delta \int_Q \frac{\sum_a s_ad_a(\vec{q})d_a(\vec{q}-\vec{p})}{\mbox{det}_M^Q\mbox{det}_M^{P-Q}}\\
  \nonumber &+8\bar{g}^2\delta\int_Q \frac{\sum_as_ad^2_a(\vec{q})}{(\mbox{det}_M^Q)^2\mbox{det}_M^{P-Q}}\mbox{tr}(\mathcal{M}_\psi^{-Q}\mathcal{M}_\psi^{Q-P}).
\end{align}

We first discuss $\Delta \Sigma_\psi(P)$. By expanding the boson propagator in powers of $\rmi p_0$ and $p^2$ according to Eq. (\ref{rg19b}) we obtain
\begin{align}
 \nonumber \Delta \Sigma_\psi(P) &\simeq 4\bar{g}^2\delta \int_Q \frac{1}{\mathcal{P}_\phi^Q\mbox{det}_M^Q}\Bigl(-\frac{\rmi \bar{y}p_0+p^2}{\mathcal{P}_\phi^Q}+\frac{4(\vec{q}\cdot\vec{p})^2}{(\mathcal{P}_\phi^Q)^2}\Bigr)\\
 \label{delta33} &\times \Biggl[\sum_as_ad_a(\vec{q})\gamma_a-\frac{2}{\mbox{det}_M^Q}\Bigl(\sum_a s_a d^2_a(\vec{q})\Bigr)\mathcal{M}_\psi^{-Q}\Biggr].
\end{align}
We neglect the $P$-independent contribution here. The frequency integration yields
\begin{align}
 \nonumber \Delta\Sigma_\psi(P) &\simeq 4\bar{g}^2\delta\int_{\vec{q}}^\prime \Biggl[ \Bigl(\sum_as_ad_a(\vec{q})\gamma_a\Bigr)\Bigl(\frac{C_1}{q^6}+\frac{C_2(\vec{q}\cdot\vec{p})^2}{q^8}\Bigr)\\
 \nonumber &+\Bigl(\sum_a s_ad_a^2(\vec{q})\Bigr)\Bigl(\frac{C_3}{q^8}\mathbb{1}_N+\frac{C_4}{q^{10}}d_b(\vec{q})\gamma_b\\
 \label{delta34} &+\frac{C_5}{q^{10}}(\vec{q}\cdot\vec{p})^2 \mathbb{1}_N+\frac{C_6(\vec{q}\cdot\vec{p})^2}{q^{12}}d_b(\vec{q})\gamma_b\Bigr)\Biggr]
\end{align}
with
\begin{align}
C_1 &= -\frac{S(\rmi \bar{y}p_0+p^2)}{2[S+(1-\bar{x})\bar{y}]^2},\ C_2 = \frac{2S^2}{[S+(1-\bar{x})\bar{y}]^3},\\
 C_3 &= -\frac{S\bar{y}(\rmi \bar{y}p_0+p^2)}{[S+(1-\bar{x})\bar{y}]^3},\\
 C_4 &= \frac{S(\rmi \bar{y} p_0+p^2)[S-(\bar{x}-3)\bar{y}]}{2[S+(1-\bar{x})\bar{y}]^3},\\
 C_5 &= \frac{6S^2 \bar{y}}{[S+(1-\bar{x})\bar{y}]^4},\ C_6 = -\frac{2S^2[S-(\bar{x}-4)\bar{y}]}{[S+(1-\bar{x})\bar{y}]^4}.
\end{align}
We observe that several $p_0$- and $p$-dependencies are generated. To further evaluate this expression we write $(\vec{q}\cdot\vec{p})^2$ according to Eq. (\ref{rg19e}), and
use the results of App. (\ref{AppdTech}) with $d=4$ to express integrals of $d$-functions in terms of
\begin{align}
J_{ab\dots c}=\mbox{tr}(\Lambda^a\Lambda^b\dots\Lambda^c).
\end{align}
In particular, for fixed $a$ we have
\begin{align}
 \label{delta36} &\int_{\vec{q}}f(q^2) d_a(\vec{q})d_b(\vec{q})d_c(\vec{q}) = \Bigl(\frac{2}{3}\Bigr)^{3/2}\frac{1}{24} J_{abc}\int_{\vec{q}}f(q^2)q^6,\\
 \nonumber &\int_{\vec{q}}f(q^2) d_a^2(\vec{q})d_b(\vec{q})d_c(\vec{q}) = \\
 \label{delta37} &\hspace{5mm} = \frac{1}{270} \Bigl[\delta_{bc}+2\delta_{ab}\delta_{ac}+2J_{aabc}+J_{abac}\Bigr]\int_{\vec{q}}f(q^2)q^8.
\end{align}
By also applying Eqs. (\ref{rg19de}) and (\ref{rg19df}) we are left with
\begin{align}
 \nonumber &\Delta \Sigma_\psi(P) \simeq 4\bar{g}^2\delta\ \Biggl[ \frac{C_2}{12} \Bigl(\sum_a s_a d_a(\vec{p})\gamma_a\Bigr) +\frac{C_3}{9}(\sum_a s_a)\mathbb{1}_N\\
 \nonumber &+\bar{C}_4\Bigl(\sum_{a,b} s_a J_{aab} \gamma_b\Bigr)+\frac{3}{4}\bar{C}_5\Bigl(\sum_{a,b}s_aJ_{aab} d_b(\vec{p})\Bigr)\mathbb{1}_N\\
 \nonumber &+\frac{C_6}{360} \sum_{a,b,c} s_a \Bigl[\delta_{bc}+2\delta_{ab}\delta_{ac}+2J_{aabc}+J_{abac}\Bigr]d_c(\vec{p})\gamma_b \\
 \label{delta38}  &+\frac{C_5}{36}p^2(\sum_a s_a)\mathbb{1}_N+\frac{\bar{C}_6}{4}p^2\Bigl(\sum_{a,b} s_a J_{aab} \gamma_b\Bigr)\Biggr]\times \int_{\vec{q}}^\prime \frac{1}{q^4},
\end{align}
where $\bar{C}_i=(\frac{2}{3})^{3/2}\frac{1}{24}C_i$. In App. \ref{AppGell} we show that
\begin{align}
 \label{delta8} \sum_a s_a J_{aab} &=0,\\
 \label{delta9} \sum_a s_a J_{aabc} &= \frac{2(d-1)(d-2)}{d}\delta_{bc},\\
 \label{delta10} \sum_a s_a J_{abac} &= 2\Bigl(2s_b-\frac{d-2}{d}\Bigr) \delta_{bc}.
\end{align}
Hence the terms containing $J_{aab}$ vanish. We further use $\sum_a 1=9$ and $\sum_a s_a = 3$ for $d=4$. This yields
\begin{align}
 \nonumber &\Delta \Sigma_\psi(P) \simeq 4\bar{g}^2\delta\ \Bigl[\frac{1}{12}\Bigl(C_2+\frac{C_6}{5}\Bigr)\Bigl(\sum_a s_ad_a(\vec{p})\gamma_a\Bigr)\\
 \label{delta39} &\hspace{2mm} +\frac{1}{3}\Bigl(C_3+\frac{C_5}{4}p^2\Bigr)\mathbb{1}_N +\frac{C_6}{45} \Bigl(\sum_a d_a(\vec{p})\gamma_a\Bigr)\Bigr] \int_{\vec{q}}^\prime \frac{1}{q^4}.
\end{align}
The contributions to $\bar{S}$, $A_\psi$, $\bar{x}$, and $\bar{\delta}$ can now be read off from the terms multiplying $\rmi p_0 \mathbb{1}_N$, $d_a(\vec{p})\gamma_a$, $p^2 \mathbb{1}_N$, and $s_ad_a(\vec{p})\gamma_a$ in this expression. We arrive at
\begin{align}
  \label{delta44} \Delta z &= -\frac{4}{3} \frac{y^2}{[1+(1-x)y]^3} g^2\delta,\\
 \label{delta45} \Delta \eta_\psi &= -\frac{8}{45}\frac{1-(x-4)y}{[1+(1-x)y]^4}g^2\delta,\\
 \label{delta46} \Delta \dot{x} &= \frac{2}{3} \frac{y[1-2(1-x)y]}{[1+(1-x)y]^4}g^2\delta,\\
 \label{delta47} \dot{\delta} &= -\eta_\psi \delta +\frac{8}{15} \frac{1+(1/4-x)y}{[1+(1-x)y]^4}g^2\delta
\end{align}
to order $O(\delta^2)$. By neglecting terms of orders $O(y\delta)$ and $O(x\delta)$ in Eq. (\ref{delta39}) we indeed recover Eq. (\ref{delta2}). Note that $z$ and $\dot{x}$ also acquire an implicit dependence on $\delta$ due to terms linear in $\eta_\psi$.

We now turn to the evaluation of $\Delta \Sigma_\phi(P)$ given in Eq. (\ref{delta32}). For the frequency dependence we have
\begin{align}
 \nonumber \frac{\partial \Delta \Sigma_\phi}{\partial \rmi p_0}\Bigr|_{P=0} &\simeq 4N\bar{g}^2\delta \frac{\bar{x}(3+\bar{x}^2)}{4(1-\bar{x}^2)^3} \int_{\vec{q}}^\prime \sum_a s_a d_a^2(\vec{q})\frac{1}{q^8}\\
 \label{delta47b} &=\frac{N\bar{g}^2\bar{x}\delta}{(1-\bar{x}^2)^3}\Bigl(1+\frac{\bar{x}^2}{3}\Bigr)\int_{\vec{q}}^\prime \frac{1}{q^4}.
\end{align}
Next we consider the $p^2$-dependence of $\Sigma_\phi$ in Eq. (\ref{delta32}). The second term therein is conceptually analogous to the case of $\delta=0$ discussed in Eqs. (\ref{rg19h})-(\ref{rg19j}). More involved, however, is the first term as due to the appearance of $s_a$ in the sum we cannot apply $\sum_a d_a(\vec{q})d_a(\vec{q}-\vec{p})= \frac{1}{3}\{4[\vec{q}\cdot(\vec{q}-\vec{p})]^2-q^2(\vec{q}-\vec{p})^2\}$ to remove the external $p$ from the $d$-functions. Instead, we write
\begin{align}
 \label{delta48} d_a(\vec{q}-\vec{p}) = d_a(\vec{q}) -2\sqrt{\frac{d}{2(d-1)}}q_i(\Lambda^a)_{ij}p_j+d_a(\vec{p})
\end{align}
in $\sum_a d_a(\vec{q})d_a(\vec{q}-\vec{p})$. We then replace $\vec{p}\to t \vec{p}$ in $\Delta \Sigma_\phi(P)$, which implies $d_a(\vec{p}) \to t^2d_a(\vec{p})$. Analogous to Eq. (\ref{rg19i}) we compute
\begin{align}
 \nonumber \Delta I_t &=\frac{1}{2}\frac{\partial^2\Delta \Sigma_\phi}{\partial t^2}\Bigr|_{p_0=t=0} \simeq -\frac{N\bar{g}^2\delta}{2S(1-\bar{x}^2)}\int_{\vec{q}}^\prime \sum_a s_ad_a(\vec{q})\\
 \nonumber &\times\Biggl[(-3q^2p^2+14(\vec{q}\cdot\vec{p})^2)d_a(\vec{q})\\
 \nonumber & -12\sqrt{\frac{d}{2(d-1)}}q^2(\vec{q}\cdot\vec{p})q_i\Lambda^a_{ij}p_j+ 2q^4d_a(\vec{p})\Biggr] \frac{1}{q^{10}}\\
 \nonumber &+  \frac{N\bar{g}^2\delta}{6S(1-\bar{x}^2)^2} \int_{\vec{q}}^\prime \Bigl(\sum_a s_ad^2_a(\vec{q})\Bigr)\\
 \label{delta49} &\times \frac{p^2q^2(\bar{x}^2-21)+4(9+2\bar{x}^2)(\vec{q}\cdot\vec{p})^2}{q^{10}}.
\end{align}
We observe the appearance of the term
\begin{align}
 \nonumber &\sqrt{\frac{d}{2(d-1)}} \int_{\vec{q}} f(q^2) d_a(\vec{q})(\vec{q}\cdot\vec{p})q_i(\Lambda^b)_{ij}p_j\\
 \nonumber &\hspace{10mm}= \frac{d}{2(d-1)} p_jp_k (\Lambda^a)_{mn}(\Lambda^b)_{ij} \int_{\vec{q}}f(q^2)q_iq_kq_mq_n\\
 \label{delta50} &\hspace{10mm}=\frac{1}{(d-1)(d+2)}p_i(\Lambda^a\Lambda^b)_{ij}p_j \int_{\vec{q}} f(q^2) q^4
\end{align}
with $a=b$. As a consequence of Eq. (\ref{delta15}) discussed below we have
\begin{align}
 \label{delta51} \sum_a s_a  p_i (\Lambda^a\Lambda^a)_{ij} p_j = \frac{(d-1)(d-2)}{d}p^2.
\end{align}
We arrive at
\begin{align}
 \nonumber \Delta I_t  &\simeq \frac{5}{12}\frac{N \bar{g}^2\delta}{S(1-\bar{x}^2)}  p^2\int_{\vec{q}}^\prime \frac{1}{q^4}\\
\label{delta52}  &- \frac{2}{3}\frac{N\bar{g}^2\delta}{S(1-\bar{x}^2)^2}\Bigl(1-\frac{\bar{x}^2}{4}\Bigr)p^2\int_{\vec{q}}^\prime \frac{1}{q^4}.
\end{align}
We employed again Eq. (\ref{delta8}). Accordingly, we have
\begin{align}
  \label{delta53} \frac{\partial \Delta \Sigma_\phi}{\partial p^2}\Bigr|_{P=0} = \frac{\partial \Delta I_t}{\partial p^2} = -\frac{N\bar{g}^2\delta(1+\bar{x}^2)}{4S(1-\bar{x}^2)^2} \int_{\vec{q}}^\prime \frac{1}{q^4}
\end{align}
and
\begin{align}
 \label{delta56} \Delta \dot{y} &= \frac{Nx}{(1-x^2)^3}\Bigl(1+\frac{x^2}{3}\Bigr)  g^2\delta,\\
 \label{delta57} \Delta \eta_\phi &=-\frac{N}{4} \frac{1+x^2}{(1-x^2)^2}g^2\delta
\end{align}
to order $O(\delta^2)$.

\section{Computational tool box}

In the following we derive a set of non-trivial relations for the $d$-functions $d_a(\vec{p})$ and the generalized real $d\times d$ Gell-Mann matrices $\Lambda^a$. We define the $d$-functions by
\begin{align}
 \label{d2} d_a(\vec{p}) = \sqrt{ \frac{d}{2 (d-1)} } p_i (\Lambda^a)_{ij} p_j,
\end{align}
where $\vec{p}=(p_1,\dots,p_d)^{\rm T}$. Furthermore, for the $\Lambda^a$ we employ the same conventions as presented in Appendix A of Ref. \cite{janssen}. Therein, explicit expressions for the matrices in 2D, 3D, and 4D can also be found.

The matrices $\Lambda^a$ are symmetric, traceless, orthogonal according to
\begin{align}
\mbox{tr}(\Lambda^a\Lambda^b)=2\delta_{ab},
\end{align}
and there are $\frac{d(d-1)}{2}+(d-1)=\frac{(d-1)(d+2)}{2}$ of them. Every real symmetric $d\times d$-matrix  $T$ can be written as
\begin{align}
 \label{d2b} T = \frac{1}{d} \mbox{tr}(T) \mathbb{1}_d + \frac{1}{2}\mbox{tr}(T \Lambda^a) \Lambda^a.
\end{align}
This implies the $\Lambda^a$ to satisfy
\begin{align}
 \label{d3} (\Lambda^a)_{ij} (\Lambda^a)_{lm} = \delta_{il}\delta_{jm} + \delta_{im}\delta_{jl}-\frac{2}{d} \delta_{ij}\delta_{lm}.
\end{align}
As a consequence we have
\begin{align}
 \label{d4} \sum_a d_a(\vec{p})d_a(\vec{k}) = \frac{1}{d-1} \Bigl( d (\vec{p}\cdot\vec{k})^2 - p^2k^2\Bigr).
\end{align}
This constitutes the addition theorem of $\ell=2$ spherical harmonics in $d$ dimensions. For $\vec{p}=\vec{k}$ we obtain
\begin{align}
 \label{d5} \sum_a d_a^2(\vec{p}) = p^4.
\end{align}
From the five $\Lambda^a$ in $d=3$ dimensions we recover the expressions in Eq. (\ref{qbt2b}). For $d=4$ we have nine distinct $d$-functions.

\subsection{$d$-Function technology}\label{AppdTech}
The computation of the RG beta functions requires the evaluation of integrals of the type
\begin{align}
 \label{d1} \int_{\vec{q}} f(q^2) d_{a_1}(\vec{q})\dots d_{a_n}(\vec{q}),
\end{align}
where $\int_{\vec{q}}=\int \mbox{d}^d q$ and $f(q^2)$ is a function of $q^2=\vec{q}^2$ which decays sufficiently fast for large momenta such that the integral in Eq. (\ref{d1}) is finite. For instance, $f(q^2)$ may have finite support in the interval $q\in[\Lambda/b,\Lambda]$ as is the case in the derivation of the RG beta functions. In this section we give a closed formula for the computation of such integrals.

We start by discussing the cases $n=1,2$. For $n=1$ we have
\begin{align}
 \label{d11} \int_{\vec{q}} f(q^2) d_a(\vec{q}) = \Bigl(\frac{d}{2(d-1)}\Bigr)^{1/2}  \int_{\vec{q}} f(q^2) q_i (\Lambda^a)_{ij} q_j.
\end{align}
The integrand is odd except for the terms with $i=j$. We can thus replace $q_iq_j=q^2\delta_{ij}$ and arrive at
\begin{align}
 \label{d12} \int_{\vec{q}} f(q^2) d_a(\vec{q})  = \Bigl(\frac{d}{2(d-1)}\Bigr)^{1/2}  \int_{\vec{q}} f(q^2) q^2 (\Lambda^a)_{ii} =0
\end{align}
due to $\mbox{tr}(\Lambda^a)=0$. For $n=2$ we have
\begin{align}
 \label{d13} \int_{\vec{q}} f(q^2) d_a(\vec{q})d_b(\vec{q}) &= \Bigl(\frac{d}{2(d-1)}\Bigr) \\
 \nonumber &\times \int_{\vec{q}} f(q^2) q_i q_jq_m q_n (\Lambda^a)_{ij} (\Lambda^b)_{mn}.
\end{align}
Again we have to build a symmetric tensor out of the $q_i$. We apply the ansatz
\begin{align}
 \label{d14} q_i q_j q_m q_n = A q^4 (\delta_{ij}\delta_{mn}+\delta_{im}\delta_{jn}+\delta_{in}\delta_{jm})
\end{align}
and contract the pairs $(i,j)$ and $(m,n)$. This leaves us with $A = \frac{1}{d(d+2)}$. We conclude that
\begin{align}
 \nonumber \int_{\vec{q}} f(q^2) d_a(\vec{q})d_b(\vec{q}) &= \Bigl(\frac{d}{2(d-1)}\Bigr) \frac{1}{d(d+2)} \int_{\vec{q}} f(q^2) q^4\\
 \nonumber  &\hspace{-15mm} \times  (\delta_{ij}\delta_{mn}+\delta_{im}\delta_{jn}+\delta_{in}\delta_{jm}) (\Lambda^a)_{ij} (\Lambda^b)_{mn}\\
 \nonumber &\hspace{-15mm}= \Bigl(\frac{d}{2(d-1)}\Bigr) \frac{2}{d(d+2)} \int_{\vec{q}} f(q^2) q^4\mbox{tr}(\Lambda^a\Lambda^b)\\
 \label{d15} &\hspace{-15mm}= \frac{1}{(d-1)(d+2)} J_{ab} \int_{\vec{q}} f(q^2) q^4,
\end{align}
where we used again that the $\Lambda^a$ are symmetric and traceless, and introduced the structure constants
\begin{align}
  \label{d16} J_{ab\dots c} =\mbox{tr}(\Lambda^a \Lambda^b \dots \Lambda^c).
\end{align}
Due to $\mbox{tr}(\Lambda^a)=0$ and $\mbox{tr}(\Lambda^a\Lambda^b)=2\delta_{ab}$ we have
\begin{align}
 J_a=0,\ J_{ab}=2\delta_{ab}.
\end{align}

The outlined procedure for computing the integrals for $n=1,2$ also applies to the cases $n>2$: Rewrite the integrand in terms of the Gell-Mann matrices and symmetrize it such that it is even under sign changes of the $q_i$. The remaining integrand is $f(q^2) q^{2n}$ times a prefactor depending on $d$, $n$, and the structure constants $J_{ab\dots c}$.  In the following we give an alternative route to evaluate the integral.

We show that for any function $f(q^2)$ we have
\begin{align}
 \nonumber &\int_{\vec{q}} f(q^2) d_{a_1}(\vec{q}) \dots d_{a_n}(\vec{q}) \\
 \label{d19} &=\Bigl(\frac{d}{2(d-1)}\Bigr)^{n/2} (-1)^n T_{a_1\dots a_n} \frac{1}{C_{d,n}} \int_{\vec{q}}f(q^2)q^{2n},
\end{align}
where the tensor $T_{a_1\dots a_n}$ is constructed from the matrix
\begin{align}
 \label{d20} M = \beta_0 \mathbb{1}_d + \alpha_a \Lambda^a
\end{align}
according to
\begin{align}
 \label{d21} T_{a_1\dots a_n} = \Bigl(\partial_{\alpha_1} \dots \partial_{\alpha_n} \frac{1}{\sqrt{\mbox{det}M}}\Bigr)_{\beta_0=1,\vec{\alpha}=0}.
\end{align}
We abbreviate $\partial_{\alpha_n}=\partial/\partial \alpha_{a_n}$. The constants $C_{d,n}$ for $n\geq 1$  are given by
\begin{align}
 \label{d22} C_{d,n} = \frac{d}{2}\Bigl(\frac{d}{2}+1\Bigr)\dots\Bigl(\frac{d}{2}+n-1\Bigr)
\end{align}
and $C_{d,0}=1$. Thus we have $C_{d,n}=(\frac{d}{2})_{n}$ with (rising) Pochhammer symbol $(a)_n$.

We first discuss some preliminaries. For a real and positive $d \times d$ matrix $M$ and real coordinates $y_i$ we have
\begin{align}
 \label{d23} \int \mbox{d}^d y\ e^{-y_i M_{ij}y_j} = \frac{\pi^{d/2}}{\sqrt{\mbox{det}M}}.
\end{align}
In particular, this implies
\begin{align}
 \label{d24} \int \mbox{d}^dy\ e^{-\beta_0 y^2} =\frac{\pi^{d/2}}{\beta_0^{d/2}},
\end{align}
and
\begin{align}
 \label{d25} \int \mbox{d}^dy \ y^{2n} e^{-\beta_0y^2}&=  (-\partial_{\beta_0})^n  \frac{\pi^{d/2}}{\beta_0^{d/2}} = C_{n,d}\frac{\pi^{d/2}}{\beta_0^{d/2+n}}.
\end{align}
Note also that with the above choice of $M=\beta_0\mathbb{1}_d +\vec{\alpha}\cdot\vec{\Lambda}$ we have
\begin{align}
 \nonumber &\Bigl(\partial_{\alpha_1}\dots\partial_{\alpha_n} \int \mbox{d}^d y \ e^{-y_iM_{ij}y_j} \Bigr)_{\vec{\alpha}=0} \\
 \label{d26} &= \Bigl(\frac{2(d-1)}{d}\Bigr)^{n/2}(-1)^n \int \mbox{d}^dy\ d_{a_1}(\vec{y})\dots d_{a_n}(\vec{y}) e^{-\beta_0 y^2}.
\end{align}
Furthermore, for arbitrary $\beta_0\neq 0$  we have
\begin{align}
 \nonumber &\Bigl(\partial_{\alpha_1} \dots \partial_{\alpha_n} \frac{1}{\sqrt{\mbox{det}M}}\Bigr)_{\vec{\alpha}=0} \\
 \nonumber &\hspace{10mm} =\frac{1}{\beta_0^{d/2+n}}\Bigl(\partial_{\alpha_1} \dots \partial_{\alpha_n}  \frac{1}{\sqrt{\mbox{det}M}}\Bigr)_{\beta_0=1,\vec{\alpha}=0} \\
 \label{d27} &\hspace{10mm} = \frac{1}{\beta_0^{d/2+n}} T_{a_1\dots a_n},
\end{align}
which might be seen most easily from the integral representation in Eq. (\ref{d23}) by introducing the variable $\hat{y}_i=\sqrt{\beta_0}y_i$.

Now we prove formula (\ref{d19}). We define the function $\bar{f}(q^2)$ by
\begin{align}
 \label{d28} f(q^2) = \bar{f}(q^2) e^{-\beta_0 q^2}
\end{align}
and find
\begin{align}
 \nonumber & \int \mbox{d}^dq\ d_{a_1}(\vec{q})\dots d_{a_n}(\vec{q}) f(q^2) \\
 \nonumber &=  \int \mbox{d}^dq\ d_{a_1}(\vec{q})\dots d_{a_n}(\vec{q}) \bar{f}(q^2) e^{-\beta_0 q^2}\\
  \nonumber &= \bar{f}(-\partial_{\beta_0}) \int \mbox{d}^dq\ d_{a_1}(\vec{q})\dots d_{a_n}(\vec{q}) e^{-\beta_0 q^2}\\
  \nonumber &=\bar{f}(-\partial_{\beta_0})\Bigl(\frac{d}{2(d-1)}\Bigr)^{n/2} (-1)^n \\
 \nonumber &\times  \Bigl(\partial_{\alpha_1}\dots\partial_{\alpha_n}\int\mbox{d}^dq\ e^{-q_iM_{ij}q_j}\Bigr)_{\vec{\alpha}=0}\\
 \nonumber &=\bar{f}(-\partial_{\beta_0})\Bigl(\frac{d}{2(d-1)}\Bigr)^{n/2} (-1)^n \frac{\pi^{d/2}}{\beta_0^{d/2+n}} T_{a_1\dots a_n}\\
 \nonumber &= \Bigl(\frac{d}{2(d-1)}\Bigr)^{n/2} (-1)^n T_{a_1\dots a_n} \\
 \nonumber &\times \bar{f}(-\partial_{\beta_0}) \frac{1}{C_{d,n}} \int \mbox{d}^dq\ q^{2n} e^{-\beta_0 q^2}\\
 \label{d29} &=  \Bigl(\frac{d}{2(d-1)}\Bigr)^{n/2} (-1)^n T_{a_1\dots a_n}  \frac{1}{C_{d,n}} \int \mbox{d}^dq\ f(q^2) q^{2n}.
\end{align}
This completes the proof.

Let us apply Eq. (\ref{d19}) for $n\leq 4$. To obtain the tensors $T_{a_1\dots a_n}$ we iteratively apply the differentiation rule
\begin{align}
 \label{d30} \frac{\mbox{d}}{\mbox{d}\alpha_i} \mbox{det}(M) = \mbox{det}(M) \cdot \mbox{tr}\Bigl(M^{-1}\frac{\mbox{d}}{\mbox{d}\alpha_i} M\Bigr)
\end{align}
with $M = \mathbb{1}_d + \alpha_a \Lambda^a$. We find
\begin{align}
 \label{d30b} &T_a =0,\ T_{ba} = \delta_{ab},\  T_{cba} =-J_{abc},\\
 \nonumber &T_{ecba} = \delta_{ab}\delta_{ce}+\delta_{ac}\delta_{be}+\delta_{ae}\delta_{bc}+J_{abce}+J_{bace}+J_{bcae},
\end{align}
and
\begin{align}
  \nonumber C_{d,1} &= \frac{d}{2},\ C_{d,2} = \frac{d}{2}\Bigl(\frac{d}{2}+1\Bigr),\  C_{ d,3} = \frac{d}{2}\Bigl(\frac{d}{2}+1\Bigr)\Bigl(\frac{d}{2}+2\Bigr),\\
  \label{d30c}  C_{ d,4} &= \frac{d}{2}\Bigl(\frac{d}{2}+1\Bigr)\Bigl(\frac{d}{2}+2\Bigr)\Bigl(\frac{d}{2}+3\Bigr).
\end{align}
We then obtain Eqs. (\ref{d12}) and (\ref{d15}) for $n=1$ and $n=2$, together with
\begin{align}
 \nonumber \int_{\vec{q}} f(q^2) d_a(\vec{q})d_b(\vec{q})d_c(\vec{q}) &= \Bigl(\frac{d}{2(d-1)}\Bigr)^{3/2}J_{abc} \\
 \label{d17} &\times \frac{8}{d(d+2)(d+4)}  \int_{\vec{q}} f(q^2) q^6
\end{align}
and
\begin{align}
  \nonumber &\int_{\vec{q}}f(q^2) d_a(\vec{q})d_b(\vec{q})d_c(\vec{q})d_e(\vec{q}) = \Bigl(\frac{d}{2(d-1)}\Bigr)^{2}\\
 \nonumber &\hspace{5mm} \times \frac{16}{d(d+2)(d+4)(d+6)} \int_{\vec{q}} f(q^2)q^8\\
 \label{d18} &\hspace{5mm} \times \Bigl[\delta_{ab}\delta_{ce} +\delta_{ac}\delta_{be}+ \delta_{ae}\delta_{bc} + J_{abce}+ J_{bace}  +  J_{bcae}\Bigr].
\end{align}
The latter two formulas can easily be verified by computing the integral with the symmetrization procedure described above.

\subsection{Some sums of real Gell-Mann matrices}\label{AppGell}

In this section we proof Eqs. (\ref{delta8})-(\ref{delta10}). For this purpose we denote the square of $\Lambda^a$ by
\begin{align}
 \label{delta11} D^a =(\Lambda^a)^2.
\end{align}
We claim that $D^a$ is a diagonal matrix. This is obvious for the diagonal $\Lambda^a$s. For an off-diagonal $\Lambda^a$ we use the bra-ket notation of Ref. \cite{janssen} and write $\Lambda^a = |j\rangle\langle k| +|k\rangle\langle j|$ with $1\leq j,k \leq d$ and $j\neq k$. We then find $(\Lambda^a)^2=|j\rangle\langle j| +|k\rangle\langle k|$, which is diagonal. In fact, for an off-diagonal $\Lambda^a$, $D^a$ is zero except for two entries of unity along the diagonal. The positions of these unities along the diagonal are uniformly distributed among the off-diagonal $\Lambda^a$s. Thus when summing up all $d(d-1)/2$ ones of them, each contributes two unities to some of the $d$ diagonal elements and we arrive at
\begin{align}
 \label{delta12} \sum_{a,\rm off-diag} D^a = \Bigl(\frac{d(d-1)}{2} \times 2 \times \frac{1}{d}\Bigr)\mathbb{1}_d = (d-1)\mathbb{1}_d.
\end{align}
Furthermore, contracting the pair of indices $(j,l)$ in Eq. (\ref{d3}) we arrive at
\begin{align}
 \label{delta13} \sum_a D^a = \frac{(d-1)(d+2)}{d}\mathbb{1}_d.
\end{align}
By subtracting the last two equations we obtain
\begin{align}
 \label{delta14} \sum_{a,\rm diag} D^a = \frac{2(d-1)}{d}\mathbb{1}_d.
\end{align}
This expression may also be obtained using the explicit expression of Eq. (A1) in Ref. \cite{janssen} for the diagonal $\Lambda^a$s. Using these results we eventually find
\begin{align}
 \label{delta15} \sum_a s_a D^a = \frac{(d-1)(d-2)}{d}\mathbb{1}_d.
\end{align}
We now easily verify
\begin{align}
 \nonumber \sum_a s_a J_{aab} &= \sum_a s_a \mbox{tr}(D^a \Lambda^b) = \mbox{tr}\Bigl(\sum_a s_aD^a \Lambda^b\Bigr)\\
 \label{delta16} &= \frac{(d-1)(d-2)}{d} \mbox{tr}(\Lambda^b) = 0,\\
 \nonumber \sum_a s_a J_{aabc} &= \frac{(d-1)(d-2)}{d} \mbox{tr}(\Lambda^b\Lambda^c) \\
 \label{delta17} &= \frac{2(d-1)(d-2)}{d}\delta_{bc}.
\end{align}
Note that with an analogous computation we obtain the sums $\sum_a J_{aab}=0$ and $\sum_a J_{aabc}$.

For the proof of Eq. (\ref{delta10}) we introduce the matrix
\begin{align}
 \label{delta18} E^{ab} = \Lambda^a \Lambda^b \Lambda^a,
\end{align}
where $a$ is not summed over. For a reason that will become clear below we are interest in computing $\sum_a s_a E^{ab}$. For this we apply a similar strategy that lead to the $a$-sums of $D^a$. First we employ Eq. (\ref{d5}) to find
\begin{align}
 \label{delta19} \sum_a (\Lambda^a\Lambda^b\Lambda^a)_{im} &= (\Lambda^b)_{jl} \sum_a (\Lambda^a)_{ij}(\Lambda^a)_{lm} = \frac{d-2}{d}(\Lambda^b)_{im}.
\end{align}
Hence
\begin{align}
\label{delta20} \sum_a E^{ab} = \frac{d-2}{d} \Lambda^b.
\end{align}
Furthermore, for every fixed $b$ we have
\begin{align}
 \label{delta21} \sum_{a,\rm off-diag} E^{ab} = s_b \Lambda^b.
\end{align}
This can be verified by using the representations $|j\rangle\langle k|+|k\rangle \langle j|$ with $j\neq k$ for off-diagonal $\Lambda^a$s and $\text{diag}(\lambda_1,\dots,\lambda_d)$ for diagonal ones. From the difference of Eqs. (\ref{delta20}) and (\ref{delta21}) we then deduce
\begin{align}
 \label{delta22} \sum_{a, \rm diag}E^{ab} = \Bigl(\frac{d-2}{d}-s_b\Bigr) \Lambda^b,
\end{align}
and, accordingly,
\begin{align}
 \label{delta23} \sum_a s_a E^{ab} = \Bigl(2s_b-\frac{d-2}{d}\Bigr)\Lambda^b.
\end{align}
This result allows for the computation of $\sum_a s_a J_{abaf}$. For this note that $\sum_a s_a E^{ab}$ is a real and symmetric matrix. In fact, it is also traceless since Eq. (\ref{delta15}) implies
\begin{align}
 \label{delta24} \mbox{tr}(\sum_a s_a E^{ab}) = \sum_a  s_a \mbox{tr}(\Lambda^a\Lambda^a \Lambda^b) = \mbox{tr}(\sum_a  s_a D^a \Lambda^b) =0.
\end{align}
Hence, $\sum_a s_a E^{ab}$ can be expanded in the basis $\{\Lambda^c\}$ via Eq. (\ref{d2b}) and we have
\begin{align}
 \label{delta25} \sum_a s_a E^{ab} = \frac{1}{2}\mbox{tr}(\sum_a s_a E^{ab} \Lambda^c) \Lambda^c = \frac{1}{2}\sum_a s_a J_{abac} \Lambda^c,
\end{align}
and, as a result of Eq. (\ref{delta23}),
\begin{align}
 \nonumber \sum_a s_a J_{abac} &= \mbox{tr}(\sum_a s_a E^{ab} \Lambda^c) = \Bigl(2s_b-\frac{d-2}{d}\Bigr) \mbox{tr}(\Lambda^b\Lambda^c)\\
 \label{delta26} &= 2\Bigl(2s_b-\frac{d-2}{d}\Bigr) \delta_{bc}.
\end{align}
Note that $\sum_a J_{abac}$ can be computed in the same manner.

\end{appendix}

\bibliographystyle{apsrev4-1}
\bibliography{refs_qbt}

\begin{thebibliography}{45}%
\makeatletter
\providecommand \@ifxundefined [1]{%
 \@ifx{#1\undefined}
}%
\providecommand \@ifnum [1]{%
 \ifnum #1\expandafter \@firstoftwo
 \else \expandafter \@secondoftwo
 \fi
}%
\providecommand \@ifx [1]{%
 \ifx #1\expandafter \@firstoftwo
 \else \expandafter \@secondoftwo
 \fi
}%
\providecommand \natexlab [1]{#1}%
\providecommand \enquote  [1]{``#1''}%
\providecommand \bibnamefont  [1]{#1}%
\providecommand \bibfnamefont [1]{#1}%
\providecommand \citenamefont [1]{#1}%
\providecommand \href@noop [0]{\@secondoftwo}%
\providecommand \href [0]{\begingroup \@sanitize@url \@href}%
\providecommand \@href[1]{\@@startlink{#1}\@@href}%
\providecommand \@@href[1]{\endgroup#1\@@endlink}%
\providecommand \@sanitize@url [0]{\catcode `\\12\catcode `\$12\catcode
  `\&12\catcode `\#12\catcode `\^12\catcode `\_12\catcode `\%12\relax}%
\providecommand \@@startlink[1]{}%
\providecommand \@@endlink[0]{}%
\providecommand \url  [0]{\begingroup\@sanitize@url \@url }%
\providecommand \@url [1]{\endgroup\@href {#1}{\urlprefix }}%
\providecommand \urlprefix  [0]{URL }%
\providecommand \Eprint [0]{\href }%
\providecommand \doibase [0]{http://dx.doi.org/}%
\providecommand \selectlanguage [0]{\@gobble}%
\providecommand \bibinfo  [0]{\@secondoftwo}%
\providecommand \bibfield  [0]{\@secondoftwo}%
\providecommand \translation [1]{[#1]}%
\providecommand \BibitemOpen [0]{}%
\providecommand \bibitemStop [0]{}%
\providecommand \bibitemNoStop [0]{.\EOS\space}%
\providecommand \EOS [0]{\spacefactor3000\relax}%
\providecommand \BibitemShut  [1]{\csname bibitem#1\endcsname}%
\let\auto@bib@innerbib\@empty
%</preamble>
\bibitem [{\citenamefont {Abrikosov}(1974)}]{abrikosov}%
  \BibitemOpen
  \bibfield  {author} {\bibinfo {author} {\bibfnamefont {A.~A.}\ \bibnamefont
  {Abrikosov}},\ }\href@noop {} {\bibfield  {journal} {\bibinfo  {journal}
  {Sov. Phys. JETP}\ }\textbf {\bibinfo {volume} {39}},\ \bibinfo {pages} {709}
  (\bibinfo {year} {1974})}\BibitemShut {NoStop}%
\bibitem [{\citenamefont {Moon}\ \emph {et~al.}(2013)\citenamefont {Moon},
  \citenamefont {Xu}, \citenamefont {Kim},\ and\ \citenamefont
  {Balents}}]{moon}%
  \BibitemOpen
  \bibfield  {author} {\bibinfo {author} {\bibfnamefont {E.-G.}\ \bibnamefont
  {Moon}}, \bibinfo {author} {\bibfnamefont {C.}~\bibnamefont {Xu}}, \bibinfo
  {author} {\bibfnamefont {Y.~B.}\ \bibnamefont {Kim}}, \ and\ \bibinfo
  {author} {\bibfnamefont {L.}~\bibnamefont {Balents}},\ }\href {\doibase
  10.1103/PhysRevLett.111.206401} {\bibfield  {journal} {\bibinfo  {journal}
  {Phys. Rev. Lett.}\ }\textbf {\bibinfo {volume} {111}},\ \bibinfo {pages}
  {206401} (\bibinfo {year} {2013})}\BibitemShut {NoStop}%
\bibitem [{\citenamefont {Abrikosov}\ and\ \citenamefont
  {Beneslavski}(1971)}]{abrben}%
  \BibitemOpen
  \bibfield  {author} {\bibinfo {author} {\bibfnamefont {A.~A.}\ \bibnamefont
  {Abrikosov}}\ and\ \bibinfo {author} {\bibfnamefont {S.~D.}\ \bibnamefont
  {Beneslavski}},\ }\href@noop {} {\bibfield  {journal} {\bibinfo  {journal}
  {Sov. Phys. JETP}\ }\textbf {\bibinfo {volume} {32}},\ \bibinfo {pages} {699}
  (\bibinfo {year} {1971})}\BibitemShut {NoStop}%
\bibitem [{\citenamefont {Luttinger}(1956)}]{luttinger}%
  \BibitemOpen
  \bibfield  {author} {\bibinfo {author} {\bibfnamefont {J.~M.}\ \bibnamefont
  {Luttinger}},\ }\href {\doibase 10.1103/PhysRev.102.1030} {\bibfield
  {journal} {\bibinfo  {journal} {Phys. Rev.}\ }\textbf {\bibinfo {volume}
  {102}},\ \bibinfo {pages} {1030} (\bibinfo {year} {1956})}\BibitemShut
  {NoStop}%
\bibitem [{\citenamefont {Murakami}\ \emph {et~al.}(2004)\citenamefont
  {Murakami}, \citenamefont {Nagosa},\ and\ \citenamefont {Zhang}}]{murakami}%
  \BibitemOpen
  \bibfield  {author} {\bibinfo {author} {\bibfnamefont {S.}~\bibnamefont
  {Murakami}}, \bibinfo {author} {\bibfnamefont {N.}~\bibnamefont {Nagosa}}, \
  and\ \bibinfo {author} {\bibfnamefont {S.-C.}\ \bibnamefont {Zhang}},\ }\href
  {\doibase 10.1103/PhysRevB.69.235206} {\bibfield  {journal} {\bibinfo
  {journal} {Phys. Rev. B}\ }\textbf {\bibinfo {volume} {69}},\ \bibinfo
  {pages} {235206} (\bibinfo {year} {2004})}\BibitemShut {NoStop}%
\bibitem [{\citenamefont {Herbut}\ and\ \citenamefont
  {Janssen}(2014)}]{herbut2014}%
  \BibitemOpen
  \bibfield  {author} {\bibinfo {author} {\bibfnamefont {I.~F.}\ \bibnamefont
  {Herbut}}\ and\ \bibinfo {author} {\bibfnamefont {L.}~\bibnamefont
  {Janssen}},\ }\href {\doibase 10.1103/PhysRevLett.113.106401} {\bibfield
  {journal} {\bibinfo  {journal} {Phys. Rev. Lett.}\ }\textbf {\bibinfo
  {volume} {113}},\ \bibinfo {pages} {106401} (\bibinfo {year}
  {2014})}\BibitemShut {NoStop}%
\bibitem [{\citenamefont {Janssen}\ and\ \citenamefont
  {Herbut}(2016)}]{janssen2015}%
  \BibitemOpen
  \bibfield  {author} {\bibinfo {author} {\bibfnamefont {L.}~\bibnamefont
  {Janssen}}\ and\ \bibinfo {author} {\bibfnamefont {I.~F.}\ \bibnamefont
  {Herbut}},\ }\href {\doibase 10.1103/PhysRevB.93.165109} {\bibfield
  {journal} {\bibinfo  {journal} {Phys. Rev. B}\ }\textbf {\bibinfo {volume}
  {93}},\ \bibinfo {pages} {165109} (\bibinfo {year} {2016})}\BibitemShut
  {NoStop}%
\bibitem [{\citenamefont {Rhim}\ and\ \citenamefont {Kim}(2015)}]{rhim}%
  \BibitemOpen
  \bibfield  {author} {\bibinfo {author} {\bibfnamefont {J.-W.}\ \bibnamefont
  {Rhim}}\ and\ \bibinfo {author} {\bibfnamefont {Y.~B.}\ \bibnamefont {Kim}},\
  }\href {\doibase 10.1103/PhysRevB.91.115124} {\bibfield  {journal} {\bibinfo
  {journal} {Phys. Rev. B}\ }\textbf {\bibinfo {volume} {91}},\ \bibinfo
  {pages} {115124} (\bibinfo {year} {2015})}\BibitemShut {NoStop}%
\bibitem [{\citenamefont {{Kondo}}\ \emph {et~al.}(2015)\citenamefont
  {{Kondo}}, \citenamefont {{Nakayama}}, \citenamefont {{Chen}}, \citenamefont
  {{Ishikawa}}, \citenamefont {{Moon}}, \citenamefont {{Yamamoto}},
  \citenamefont {{Ota}}, \citenamefont {{Malaeb}}, \citenamefont {{Kanai}},
  \citenamefont {{Nakashima}}, \citenamefont {{Ishida}}, \citenamefont
  {{Yoshida}}, \citenamefont {{Yamamoto}}, \citenamefont {{Matsunami}},
  \citenamefont {{Kimura}}, \citenamefont {{Inami}}, \citenamefont {{Ono}},
  \citenamefont {{Kumigashira}}, \citenamefont {{Nakatsuji}}, \citenamefont
  {{Balents}},\ and\ \citenamefont {{Shin}}}]{kondo}%
  \BibitemOpen
  \bibfield  {author} {\bibinfo {author} {\bibfnamefont {T.}~\bibnamefont
  {{Kondo}}}, \bibinfo {author} {\bibfnamefont {M.}~\bibnamefont {{Nakayama}}},
  \bibinfo {author} {\bibfnamefont {R.}~\bibnamefont {{Chen}}}, \bibinfo
  {author} {\bibfnamefont {J.~J.}\ \bibnamefont {{Ishikawa}}}, \bibinfo
  {author} {\bibfnamefont {E.-G.}\ \bibnamefont {{Moon}}}, \bibinfo {author}
  {\bibfnamefont {T.}~\bibnamefont {{Yamamoto}}}, \bibinfo {author}
  {\bibfnamefont {Y.}~\bibnamefont {{Ota}}}, \bibinfo {author} {\bibfnamefont
  {W.}~\bibnamefont {{Malaeb}}}, \bibinfo {author} {\bibfnamefont
  {H.}~\bibnamefont {{Kanai}}}, \bibinfo {author} {\bibfnamefont
  {Y.}~\bibnamefont {{Nakashima}}}, \bibinfo {author} {\bibfnamefont
  {Y.}~\bibnamefont {{Ishida}}}, \bibinfo {author} {\bibfnamefont
  {R.}~\bibnamefont {{Yoshida}}}, \bibinfo {author} {\bibfnamefont
  {H.}~\bibnamefont {{Yamamoto}}}, \bibinfo {author} {\bibfnamefont
  {M.}~\bibnamefont {{Matsunami}}}, \bibinfo {author} {\bibfnamefont
  {S.}~\bibnamefont {{Kimura}}}, \bibinfo {author} {\bibfnamefont
  {N.}~\bibnamefont {{Inami}}}, \bibinfo {author} {\bibfnamefont
  {K.}~\bibnamefont {{Ono}}}, \bibinfo {author} {\bibfnamefont
  {H.}~\bibnamefont {{Kumigashira}}}, \bibinfo {author} {\bibfnamefont
  {S.}~\bibnamefont {{Nakatsuji}}}, \bibinfo {author} {\bibfnamefont
  {L.}~\bibnamefont {{Balents}}}, \ and\ \bibinfo {author} {\bibfnamefont
  {S.}~\bibnamefont {{Shin}}},\ }\href {\doibase 10.1038/ncomms10042}
  {\bibfield  {journal} {\bibinfo  {journal} {Nat. Commun.}\ }\textbf {\bibinfo
  {volume} {6}},\ \bibinfo {eid} {10042} (\bibinfo {year} {2015})}\BibitemShut
  {NoStop}%
\bibitem [{\citenamefont {Savary}\ \emph {et~al.}(2014)\citenamefont {Savary},
  \citenamefont {Moon},\ and\ \citenamefont {Balents}}]{savary}%
  \BibitemOpen
  \bibfield  {author} {\bibinfo {author} {\bibfnamefont {L.}~\bibnamefont
  {Savary}}, \bibinfo {author} {\bibfnamefont {E.-G.}\ \bibnamefont {Moon}}, \
  and\ \bibinfo {author} {\bibfnamefont {L.}~\bibnamefont {Balents}},\ }\href
  {\doibase 10.1103/PhysRevX.4.041027} {\bibfield  {journal} {\bibinfo
  {journal} {Phys. Rev. X}\ }\textbf {\bibinfo {volume} {4}},\ \bibinfo {pages}
  {041027} (\bibinfo {year} {2014})}\BibitemShut {NoStop}%
\bibitem [{\citenamefont {Murray}\ \emph {et~al.}(2015)\citenamefont {Murray},
  \citenamefont {Vafek},\ and\ \citenamefont {Balents}}]{murray}%
  \BibitemOpen
  \bibfield  {author} {\bibinfo {author} {\bibfnamefont {J.~M.}\ \bibnamefont
  {Murray}}, \bibinfo {author} {\bibfnamefont {O.}~\bibnamefont {Vafek}}, \
  and\ \bibinfo {author} {\bibfnamefont {L.}~\bibnamefont {Balents}},\ }\href
  {\doibase 10.1103/PhysRevB.92.035137} {\bibfield  {journal} {\bibinfo
  {journal} {Phys. Rev. B}\ }\textbf {\bibinfo {volume} {92}},\ \bibinfo
  {pages} {035137} (\bibinfo {year} {2015})}\BibitemShut {NoStop}%
\bibitem [{\citenamefont {Levy-Leblond}(1963)}]{leblond}%
  \BibitemOpen
  \bibfield  {author} {\bibinfo {author} {\bibfnamefont {J.-M.}\ \bibnamefont
  {Levy-Leblond}},\ }\href {\doibase http://dx.doi.org/10.1063/1.1724319}
  {\bibfield  {journal} {\bibinfo  {journal} {J. Math. Phys.}\ }\textbf
  {\bibinfo {volume} {4}},\ \bibinfo {pages} {776} (\bibinfo {year}
  {1963})}\BibitemShut {NoStop}%
\bibitem [{\citenamefont {Hagen}(1972)}]{hagen}%
  \BibitemOpen
  \bibfield  {author} {\bibinfo {author} {\bibfnamefont {C.~R.}\ \bibnamefont
  {Hagen}},\ }\href {\doibase 10.1103/PhysRevD.5.377} {\bibfield  {journal}
  {\bibinfo  {journal} {Phys. Rev. D}\ }\textbf {\bibinfo {volume} {5}},\
  \bibinfo {pages} {377} (\bibinfo {year} {1972})}\BibitemShut {NoStop}%
\bibitem [{\citenamefont {Herbut}(2006)}]{herbut2006}%
  \BibitemOpen
  \bibfield  {author} {\bibinfo {author} {\bibfnamefont {I.~F.}\ \bibnamefont
  {Herbut}},\ }\href {\doibase 10.1103/PhysRevLett.97.146401} {\bibfield
  {journal} {\bibinfo  {journal} {Phys. Rev. Lett.}\ }\textbf {\bibinfo
  {volume} {97}},\ \bibinfo {pages} {146401} (\bibinfo {year}
  {2006})}\BibitemShut {NoStop}%
\bibitem [{\citenamefont {Herbut}\ \emph
  {et~al.}(2009{\natexlab{a}})\citenamefont {Herbut}, \citenamefont
  {Juri\ifmmode \check{c}\else \v{c}\fi{}i\ifmmode~\acute{c}\else \'{c}\fi{}},\
  and\ \citenamefont {Roy}}]{herjurroy}%
  \BibitemOpen
  \bibfield  {author} {\bibinfo {author} {\bibfnamefont {I.~F.}\ \bibnamefont
  {Herbut}}, \bibinfo {author} {\bibfnamefont {V.}~\bibnamefont {Juri\ifmmode
  \check{c}\else \v{c}\fi{}i\ifmmode~\acute{c}\else \'{c}\fi{}}}, \ and\
  \bibinfo {author} {\bibfnamefont {B.}~\bibnamefont {Roy}},\ }\href {\doibase
  10.1103/PhysRevB.79.085116} {\bibfield  {journal} {\bibinfo  {journal} {Phys.
  Rev. B}\ }\textbf {\bibinfo {volume} {79}},\ \bibinfo {pages} {085116}
  (\bibinfo {year} {2009}{\natexlab{a}})}\BibitemShut {NoStop}%
\bibitem [{\citenamefont {Assaad}\ and\ \citenamefont {Herbut}(2013)}]{assaad}%
  \BibitemOpen
  \bibfield  {author} {\bibinfo {author} {\bibfnamefont {F.~F.}\ \bibnamefont
  {Assaad}}\ and\ \bibinfo {author} {\bibfnamefont {I.~F.}\ \bibnamefont
  {Herbut}},\ }\href {\doibase 10.1103/PhysRevX.3.031010} {\bibfield  {journal}
  {\bibinfo  {journal} {Phys. Rev. X}\ }\textbf {\bibinfo {volume} {3}},\
  \bibinfo {pages} {031010} (\bibinfo {year} {2013})}\BibitemShut {NoStop}%
\bibitem [{\citenamefont {Parisen~Toldin}\ \emph {et~al.}(2015)\citenamefont
  {Parisen~Toldin}, \citenamefont {Hohenadler}, \citenamefont {Assaad},\ and\
  \citenamefont {Herbut}}]{parisen}%
  \BibitemOpen
  \bibfield  {author} {\bibinfo {author} {\bibfnamefont {F.}~\bibnamefont
  {Parisen~Toldin}}, \bibinfo {author} {\bibfnamefont {M.}~\bibnamefont
  {Hohenadler}}, \bibinfo {author} {\bibfnamefont {F.~F.}\ \bibnamefont
  {Assaad}}, \ and\ \bibinfo {author} {\bibfnamefont {I.~F.}\ \bibnamefont
  {Herbut}},\ }\href {\doibase 10.1103/PhysRevB.91.165108} {\bibfield
  {journal} {\bibinfo  {journal} {Phys. Rev. B}\ }\textbf {\bibinfo {volume}
  {91}},\ \bibinfo {pages} {165108} (\bibinfo {year} {2015})}\BibitemShut
  {NoStop}%
\bibitem [{\citenamefont {Otsuka}\ \emph {et~al.}(2016)\citenamefont {Otsuka},
  \citenamefont {Yunoki},\ and\ \citenamefont {Sorella}}]{sorella}%
  \BibitemOpen
  \bibfield  {author} {\bibinfo {author} {\bibfnamefont {Y.}~\bibnamefont
  {Otsuka}}, \bibinfo {author} {\bibfnamefont {S.}~\bibnamefont {Yunoki}}, \
  and\ \bibinfo {author} {\bibfnamefont {S.}~\bibnamefont {Sorella}},\ }\href
  {\doibase 10.1103/PhysRevX.6.011029} {\bibfield  {journal} {\bibinfo
  {journal} {Phys. Rev. X}\ }\textbf {\bibinfo {volume} {6}},\ \bibinfo {pages}
  {011029} (\bibinfo {year} {2016})}\BibitemShut {NoStop}%
\bibitem [{\citenamefont {Janssen}\ and\ \citenamefont
  {Herbut}(2015)}]{janssen}%
  \BibitemOpen
  \bibfield  {author} {\bibinfo {author} {\bibfnamefont {L.}~\bibnamefont
  {Janssen}}\ and\ \bibinfo {author} {\bibfnamefont {I.~F.}\ \bibnamefont
  {Herbut}},\ }\href {\doibase 10.1103/PhysRevB.92.045117} {\bibfield
  {journal} {\bibinfo  {journal} {Phys. Rev. B}\ }\textbf {\bibinfo {volume}
  {92}},\ \bibinfo {pages} {045117} (\bibinfo {year} {2015})}\BibitemShut
  {NoStop}%
\bibitem [{\citenamefont {Ballentine}(1990)}]{ballentine}%
  \BibitemOpen
  \bibfield  {author} {\bibinfo {author} {\bibfnamefont {L.~E.}\ \bibnamefont
  {Ballentine}},\ }\href@noop {} {\emph {\bibinfo {title} {{Quantum
  Mechanics}}}}\ (\bibinfo  {publisher} {Prentice Hall, New Jersey},\ \bibinfo
  {year} {1990})\BibitemShut {NoStop}%
\bibitem [{\citenamefont {Herbut}(2013)}]{herbut2013}%
  \BibitemOpen
  \bibfield  {author} {\bibinfo {author} {\bibfnamefont {I.~F.}\ \bibnamefont
  {Herbut}},\ }\href {\doibase 10.1103/PhysRevD.87.085002} {\bibfield
  {journal} {\bibinfo  {journal} {Phys. Rev. D}\ }\textbf {\bibinfo {volume}
  {87}},\ \bibinfo {pages} {085002} (\bibinfo {year} {2013})}\BibitemShut
  {NoStop}%
\bibitem [{\citenamefont {Bardeen}\ \emph {et~al.}(1957)\citenamefont
  {Bardeen}, \citenamefont {Cooper},\ and\ \citenamefont {Schrieffer}}]{bcs}%
  \BibitemOpen
  \bibfield  {author} {\bibinfo {author} {\bibfnamefont {J.}~\bibnamefont
  {Bardeen}}, \bibinfo {author} {\bibfnamefont {L.~N.}\ \bibnamefont {Cooper}},
  \ and\ \bibinfo {author} {\bibfnamefont {J.~R.}\ \bibnamefont {Schrieffer}},\
  }\href {\doibase 10.1103/PhysRev.108.1175} {\bibfield  {journal} {\bibinfo
  {journal} {Phys. Rev.}\ }\textbf {\bibinfo {volume} {108}},\ \bibinfo {pages}
  {1175} (\bibinfo {year} {1957})}\BibitemShut {NoStop}%
\bibitem [{\citenamefont {Herbut}(2007)}]{herbutbook}%
  \BibitemOpen
  \bibfield  {author} {\bibinfo {author} {\bibfnamefont {I.}~\bibnamefont
  {Herbut}},\ }\href@noop {} {\emph {\bibinfo {title} {{A Modern Approach to
  Critical Phenomena}}}}\ (\bibinfo  {publisher} {Cambridge University Press,
  Cambridge, England},\ \bibinfo {year} {2007})\BibitemShut {NoStop}%
\bibitem [{\citenamefont {Herbut}(2012)}]{herbut2012}%
  \BibitemOpen
  \bibfield  {author} {\bibinfo {author} {\bibfnamefont {I.~F.}\ \bibnamefont
  {Herbut}},\ }\href {\doibase 10.1103/PhysRevB.85.085304} {\bibfield
  {journal} {\bibinfo  {journal} {Phys. Rev. B}\ }\textbf {\bibinfo {volume}
  {85}},\ \bibinfo {pages} {085304} (\bibinfo {year} {2012})}\BibitemShut
  {NoStop}%
\bibitem [{\citenamefont {Georgi}(1999)}]{georgi}%
  \BibitemOpen
  \bibfield  {author} {\bibinfo {author} {\bibfnamefont {H.}~\bibnamefont
  {Georgi}},\ }\href@noop {} {\emph {\bibinfo {title} {{Lie Algebras in
  Particle Physics}}}}\ (\bibinfo  {publisher} {Westview press},\ \bibinfo
  {year} {1999})\ \bibinfo {note} {2nd edition, ch. 21}\BibitemShut {NoStop}%
\bibitem [{\citenamefont {Nozieres}\ and\ \citenamefont
  {Schmitt-Rink}()}]{nozieres}%
  \BibitemOpen
  \bibfield  {author} {\bibinfo {author} {\bibfnamefont {P.}~\bibnamefont
  {Nozieres}}\ and\ \bibinfo {author} {\bibfnamefont {S.}~\bibnamefont
  {Schmitt-Rink}},\ }\href {\doibase 10.1007/BF00683774} {\bibfield  {journal}
  {\bibinfo  {journal} {J. Low Temp. Phys.}\ }\textbf {\bibinfo {volume}
  {59}},\ \bibinfo {pages} {195}}\BibitemShut {NoStop}%
\bibitem [{\citenamefont {Herbut}\ \emph
  {et~al.}(2009{\natexlab{b}})\citenamefont {Herbut}, \citenamefont
  {Juri\ifmmode \check{c}\else \v{c}\fi{}i\ifmmode~\acute{c}\else \'{c}\fi{}},\
  and\ \citenamefont {Vafek}}]{herjurvaf}%
  \BibitemOpen
  \bibfield  {author} {\bibinfo {author} {\bibfnamefont {I.~F.}\ \bibnamefont
  {Herbut}}, \bibinfo {author} {\bibfnamefont {V.}~\bibnamefont {Juri\ifmmode
  \check{c}\else \v{c}\fi{}i\ifmmode~\acute{c}\else \'{c}\fi{}}}, \ and\
  \bibinfo {author} {\bibfnamefont {O.}~\bibnamefont {Vafek}},\ }\href
  {\doibase 10.1103/PhysRevB.80.075432} {\bibfield  {journal} {\bibinfo
  {journal} {Phys. Rev. B}\ }\textbf {\bibinfo {volume} {80}},\ \bibinfo
  {pages} {075432} (\bibinfo {year} {2009}{\natexlab{b}})}\BibitemShut
  {NoStop}%
\bibitem [{\citenamefont {Roy}\ \emph {et~al.}(2013)\citenamefont {Roy},
  \citenamefont {Juri\ifmmode \check{c}\else \v{c}\fi{}i\ifmmode~\acute{c}\else
  \'{c}\fi{}},\ and\ \citenamefont {Herbut}}]{PhysRevB.87.041401}%
  \BibitemOpen
  \bibfield  {author} {\bibinfo {author} {\bibfnamefont {B.}~\bibnamefont
  {Roy}}, \bibinfo {author} {\bibfnamefont {V.}~\bibnamefont {Juri\ifmmode
  \check{c}\else \v{c}\fi{}i\ifmmode~\acute{c}\else \'{c}\fi{}}}, \ and\
  \bibinfo {author} {\bibfnamefont {I.~F.}\ \bibnamefont {Herbut}},\ }\href
  {\doibase 10.1103/PhysRevB.87.041401} {\bibfield  {journal} {\bibinfo
  {journal} {Phys. Rev. B}\ }\textbf {\bibinfo {volume} {87}},\ \bibinfo
  {pages} {041401} (\bibinfo {year} {2013})}\BibitemShut {NoStop}%
\bibitem [{\citenamefont {Janssen}\ and\ \citenamefont
  {Herbut}(2014)}]{janssen2014}%
  \BibitemOpen
  \bibfield  {author} {\bibinfo {author} {\bibfnamefont {L.}~\bibnamefont
  {Janssen}}\ and\ \bibinfo {author} {\bibfnamefont {I.~F.}\ \bibnamefont
  {Herbut}},\ }\href {\doibase 10.1103/PhysRevB.89.205403} {\bibfield
  {journal} {\bibinfo  {journal} {Phys. Rev. B}\ }\textbf {\bibinfo {volume}
  {89}},\ \bibinfo {pages} {205403} (\bibinfo {year} {2014})}\BibitemShut
  {NoStop}%
\bibitem [{\citenamefont {Boettcher}\ \emph {et~al.}(2014)\citenamefont
  {Boettcher}, \citenamefont {Pawlowski},\ and\ \citenamefont
  {Wetterich}}]{PhysRevA.89.053630}%
  \BibitemOpen
  \bibfield  {author} {\bibinfo {author} {\bibfnamefont {I.}~\bibnamefont
  {Boettcher}}, \bibinfo {author} {\bibfnamefont {J.~M.}\ \bibnamefont
  {Pawlowski}}, \ and\ \bibinfo {author} {\bibfnamefont {C.}~\bibnamefont
  {Wetterich}},\ }\href {\doibase 10.1103/PhysRevA.89.053630} {\bibfield
  {journal} {\bibinfo  {journal} {Phys. Rev. A}\ }\textbf {\bibinfo {volume}
  {89}},\ \bibinfo {pages} {053630} (\bibinfo {year} {2014})}\BibitemShut
  {NoStop}%
\bibitem [{\citenamefont {Wang}\ \emph {et~al.}(2014)\citenamefont {Wang},
  \citenamefont {Corboz},\ and\ \citenamefont {Troyer}}]{wang2014}%
  \BibitemOpen
  \bibfield  {author} {\bibinfo {author} {\bibfnamefont {L.}~\bibnamefont
  {Wang}}, \bibinfo {author} {\bibfnamefont {P.}~\bibnamefont {Corboz}}, \ and\
  \bibinfo {author} {\bibfnamefont {M.}~\bibnamefont {Troyer}},\ }\href
  {http://stacks.iop.org/1367-2630/16/i=10/a=103008} {\bibfield  {journal}
  {\bibinfo  {journal} {New J. Phys.}\ }\textbf {\bibinfo {volume} {16}},\
  \bibinfo {pages} {103008} (\bibinfo {year} {2014})}\BibitemShut {NoStop}%
\bibitem [{\citenamefont {Li}\ \emph {et~al.}(2015)\citenamefont {Li},
  \citenamefont {Jiang},\ and\ \citenamefont {Yao}}]{li}%
  \BibitemOpen
  \bibfield  {author} {\bibinfo {author} {\bibfnamefont {Z.-X.}\ \bibnamefont
  {Li}}, \bibinfo {author} {\bibfnamefont {Y.-F.}\ \bibnamefont {Jiang}}, \
  and\ \bibinfo {author} {\bibfnamefont {H.}~\bibnamefont {Yao}},\ }\href
  {http://stacks.iop.org/1367-2630/17/i=8/a=085003} {\bibfield  {journal}
  {\bibinfo  {journal} {New J. Phys.}\ }\textbf {\bibinfo {volume} {17}},\
  \bibinfo {pages} {085003} (\bibinfo {year} {2015})}\BibitemShut {NoStop}%
\bibitem [{\citenamefont {Wang}\ \emph {et~al.}(2015)\citenamefont {Wang},
  \citenamefont {Iazzi}, \citenamefont {Corboz},\ and\ \citenamefont
  {Troyer}}]{wang2015}%
  \BibitemOpen
  \bibfield  {author} {\bibinfo {author} {\bibfnamefont {L.}~\bibnamefont
  {Wang}}, \bibinfo {author} {\bibfnamefont {M.}~\bibnamefont {Iazzi}},
  \bibinfo {author} {\bibfnamefont {P.}~\bibnamefont {Corboz}}, \ and\ \bibinfo
  {author} {\bibfnamefont {M.}~\bibnamefont {Troyer}},\ }\href {\doibase
  10.1103/PhysRevB.91.235151} {\bibfield  {journal} {\bibinfo  {journal} {Phys.
  Rev. B}\ }\textbf {\bibinfo {volume} {91}},\ \bibinfo {pages} {235151}
  (\bibinfo {year} {2015})}\BibitemShut {NoStop}%
\bibitem [{\citenamefont {Huffman}\ and\ \citenamefont
  {Chandrasekharan}(2014)}]{huffman}%
  \BibitemOpen
  \bibfield  {author} {\bibinfo {author} {\bibfnamefont {E.~F.}\ \bibnamefont
  {Huffman}}\ and\ \bibinfo {author} {\bibfnamefont {S.}~\bibnamefont
  {Chandrasekharan}},\ }\href {\doibase 10.1103/PhysRevB.89.111101} {\bibfield
  {journal} {\bibinfo  {journal} {Phys. Rev. B}\ }\textbf {\bibinfo {volume}
  {89}},\ \bibinfo {pages} {111101} (\bibinfo {year} {2014})}\BibitemShut
  {NoStop}%
\bibitem [{\citenamefont {Nishida}\ and\ \citenamefont {Son}(2006)}]{nishida}%
  \BibitemOpen
  \bibfield  {author} {\bibinfo {author} {\bibfnamefont {Y.}~\bibnamefont
  {Nishida}}\ and\ \bibinfo {author} {\bibfnamefont {D.~T.}\ \bibnamefont
  {Son}},\ }\href {\doibase 10.1103/PhysRevLett.97.050403} {\bibfield
  {journal} {\bibinfo  {journal} {Phys. Rev. Lett.}\ }\textbf {\bibinfo
  {volume} {97}},\ \bibinfo {pages} {050403} (\bibinfo {year}
  {2006})}\BibitemShut {NoStop}%
\bibitem [{\citenamefont {Nikoli\ifmmode~\acute{c}\else \'{c}\fi{}}\ and\
  \citenamefont {Sachdev}(2007)}]{nikolic}%
  \BibitemOpen
  \bibfield  {author} {\bibinfo {author} {\bibfnamefont {P.}~\bibnamefont
  {Nikoli\ifmmode~\acute{c}\else \'{c}\fi{}}}\ and\ \bibinfo {author}
  {\bibfnamefont {S.}~\bibnamefont {Sachdev}},\ }\href {\doibase
  10.1103/PhysRevA.75.033608} {\bibfield  {journal} {\bibinfo  {journal} {Phys.
  Rev. A}\ }\textbf {\bibinfo {volume} {75}},\ \bibinfo {pages} {033608}
  (\bibinfo {year} {2007})}\BibitemShut {NoStop}%
\bibitem [{\citenamefont {Veillette}\ \emph {et~al.}(2007)\citenamefont
  {Veillette}, \citenamefont {Sheehy},\ and\ \citenamefont
  {Radzihovsky}}]{radzihovsky}%
  \BibitemOpen
  \bibfield  {author} {\bibinfo {author} {\bibfnamefont {M.~Y.}\ \bibnamefont
  {Veillette}}, \bibinfo {author} {\bibfnamefont {D.~E.}\ \bibnamefont
  {Sheehy}}, \ and\ \bibinfo {author} {\bibfnamefont {L.}~\bibnamefont
  {Radzihovsky}},\ }\href {\doibase 10.1103/PhysRevA.75.043614} {\bibfield
  {journal} {\bibinfo  {journal} {Phys. Rev. A}\ }\textbf {\bibinfo {volume}
  {75}},\ \bibinfo {pages} {043614} (\bibinfo {year} {2007})}\BibitemShut
  {NoStop}%
\bibitem [{\citenamefont {Diehl}\ \emph {et~al.}(2007)\citenamefont {Diehl},
  \citenamefont {Gies}, \citenamefont {Pawlowski},\ and\ \citenamefont
  {Wetterich}}]{diehl}%
  \BibitemOpen
  \bibfield  {author} {\bibinfo {author} {\bibfnamefont {S.}~\bibnamefont
  {Diehl}}, \bibinfo {author} {\bibfnamefont {H.}~\bibnamefont {Gies}},
  \bibinfo {author} {\bibfnamefont {J.~M.}\ \bibnamefont {Pawlowski}}, \ and\
  \bibinfo {author} {\bibfnamefont {C.}~\bibnamefont {Wetterich}},\ }\href
  {\doibase 10.1103/PhysRevA.76.021602} {\bibfield  {journal} {\bibinfo
  {journal} {Phys. Rev. A}\ }\textbf {\bibinfo {volume} {76}},\ \bibinfo
  {pages} {021602} (\bibinfo {year} {2007})}\BibitemShut {NoStop}%
\bibitem [{\citenamefont {{Zwerger, W.}}(2012)}]{Zwerger}%
  \BibitemOpen
  \bibinfo {editor} {\bibnamefont {{Zwerger, W.}}},\ ed.,\ \href@noop {} {\emph
  {\bibinfo {title} {{The BCS-BEC Crossover and the Unitary Fermi Gas}}}}\
  (\bibinfo  {publisher} {Springer, Berlin},\ \bibinfo {year}
  {2012})\BibitemShut {NoStop}%
\bibitem [{\citenamefont {Fisher}\ \emph {et~al.}(1989)\citenamefont {Fisher},
  \citenamefont {Weichman}, \citenamefont {Grinstein},\ and\ \citenamefont
  {Fisher}}]{fisher}%
  \BibitemOpen
  \bibfield  {author} {\bibinfo {author} {\bibfnamefont {M.~P.~A.}\
  \bibnamefont {Fisher}}, \bibinfo {author} {\bibfnamefont {P.~B.}\
  \bibnamefont {Weichman}}, \bibinfo {author} {\bibfnamefont {G.}~\bibnamefont
  {Grinstein}}, \ and\ \bibinfo {author} {\bibfnamefont {D.~S.}\ \bibnamefont
  {Fisher}},\ }\href {\doibase 10.1103/PhysRevB.40.546} {\bibfield  {journal}
  {\bibinfo  {journal} {Phys. Rev. B}\ }\textbf {\bibinfo {volume} {40}},\
  \bibinfo {pages} {546} (\bibinfo {year} {1989})}\BibitemShut {NoStop}%
\bibitem [{\citenamefont {Weinrib}\ and\ \citenamefont
  {Halperin}(1983)}]{weinrib}%
  \BibitemOpen
  \bibfield  {author} {\bibinfo {author} {\bibfnamefont {A.}~\bibnamefont
  {Weinrib}}\ and\ \bibinfo {author} {\bibfnamefont {B.~I.}\ \bibnamefont
  {Halperin}},\ }\href {\doibase 10.1103/PhysRevB.27.413} {\bibfield  {journal}
  {\bibinfo  {journal} {Phys. Rev. B}\ }\textbf {\bibinfo {volume} {27}},\
  \bibinfo {pages} {413} (\bibinfo {year} {1983})}\BibitemShut {NoStop}%
\bibitem [{\citenamefont {Herbut}(1998)}]{herbut1998}%
  \BibitemOpen
  \bibfield  {author} {\bibinfo {author} {\bibfnamefont {I.~F.}\ \bibnamefont
  {Herbut}},\ }\href {\doibase 10.1103/PhysRevB.57.13729} {\bibfield  {journal}
  {\bibinfo  {journal} {Phys. Rev. B}\ }\textbf {\bibinfo {volume} {57}},\
  \bibinfo {pages} {13729} (\bibinfo {year} {1998})}\BibitemShut {NoStop}%
\bibitem [{\citenamefont {Wallace}\ and\ \citenamefont {Zia}(1975)}]{wallace}%
  \BibitemOpen
  \bibfield  {author} {\bibinfo {author} {\bibfnamefont {D.}~\bibnamefont
  {Wallace}}\ and\ \bibinfo {author} {\bibfnamefont {R.}~\bibnamefont {Zia}},\
  }\href {\doibase 10.1016/0003-4916(75)90267-5} {\bibfield  {journal}
  {\bibinfo  {journal} {Ann. Phys.}\ }\textbf {\bibinfo {volume} {92}},\
  \bibinfo {pages} {142 } (\bibinfo {year} {1975})}\BibitemShut {NoStop}%
\bibitem [{\citenamefont {Herbut}\ and\ \citenamefont {Te\ifmmode
  \check{s}\else \v{s}\fi{}anovi\ifmmode~\acute{c}\else
  \'{c}\fi{}}(1996)}]{tesanovic}%
  \BibitemOpen
  \bibfield  {author} {\bibinfo {author} {\bibfnamefont {I.~F.}\ \bibnamefont
  {Herbut}}\ and\ \bibinfo {author} {\bibfnamefont {Z.}~\bibnamefont
  {Te\ifmmode \check{s}\else \v{s}\fi{}anovi\ifmmode~\acute{c}\else
  \'{c}\fi{}}},\ }\href {\doibase 10.1103/PhysRevLett.76.4588} {\bibfield
  {journal} {\bibinfo  {journal} {Phys. Rev. Lett.}\ }\textbf {\bibinfo
  {volume} {76}},\ \bibinfo {pages} {4588} (\bibinfo {year}
  {1996})}\BibitemShut {NoStop}%
\bibitem [{\citenamefont {Bergerhoff}\ \emph {et~al.}(1996)\citenamefont
  {Bergerhoff}, \citenamefont {Freire}, \citenamefont {Litim}, \citenamefont
  {Lola},\ and\ \citenamefont {Wetterich}}]{litim}%
  \BibitemOpen
  \bibfield  {author} {\bibinfo {author} {\bibfnamefont {B.}~\bibnamefont
  {Bergerhoff}}, \bibinfo {author} {\bibfnamefont {F.}~\bibnamefont {Freire}},
  \bibinfo {author} {\bibfnamefont {D.~F.}\ \bibnamefont {Litim}}, \bibinfo
  {author} {\bibfnamefont {S.}~\bibnamefont {Lola}}, \ and\ \bibinfo {author}
  {\bibfnamefont {C.}~\bibnamefont {Wetterich}},\ }\href {\doibase
  10.1103/PhysRevB.53.5734} {\bibfield  {journal} {\bibinfo  {journal} {Phys.
  Rev. B}\ }\textbf {\bibinfo {volume} {53}},\ \bibinfo {pages} {5734}
  (\bibinfo {year} {1996})}\BibitemShut {NoStop}%
\end{thebibliography}%

\end{document}